{}
\documentclass{pasa}%

\title[The Status of Multi-Dimensional Core-Collapse Supernova Models]{The Status of Multi-Dimensional Core-Collapse Supernova Models}
\author[B.~M\"uller]{B.~M\"uller$^{1,2,3}$\\
\affil{$^1$Astrophysics Research Centre, School of Mathematics and Physics, Queen's University Belfast, Belfast, BT7~1NN, United Kingdom \url{b.mueller@qub.ac.uk}}%
\affil{$^2$Monash Centre for Astrophysics, School of Physics and Astronomy, Monash University, VIC~3800, Australia}%
\affil{$^3$Joint Institute for Nuclear Astrophysics, University of Notre Dame, IN~46556, USA}
}
\jid{PASA}
\doi{10.1017/pas.\the\year.xxx}
\jyear{\the\year}

\usepackage[authoryear]{natbib}

\usepackage{txfonts}

\bibpunct{(}{)}{;}{a}{}{,}
\usepackage{aas_macros}
\usepackage{hyperref} 
\hypersetup{colorlinks,citecolor=blue,linkcolor=blue,urlcolor=blue}

\newcommand{\pd}{\partial}
\newcommand{\ud}{\mathrm{d}}

\begin{document}%
\begin{abstract}
 Numerical models of core-collapse supernova explosions powered by the
neutrino-driven mechanism have matured considerable in recent years.
Explosions at the low-mass end of the progenitor spectrum can
routinely be simulated in 1D, 2D, and 3D today and already allow us to
study supernova nucleosynthesis based on first-principle
models. Results of nucleosynthesis calculations indicate that
supernovae of the lowest masses could be important as contributors of
some lighter neutron-rich elements beyond iron. The explosion
mechanism of more massive stars is still under investigation, although
first 3D models of neutrino-driven explosions employing multi-group
neutrino transport have recently become available. Together with
earlier 2D models and more simplified 3D simulations, these have
elucidated the interplay between neutrino heating and
multi-dimensional hydrodynamic instabilities in the post-shock region
that is essential for shock revival. However, some physical
ingredients may still need to be added or improved before simulations can robustly
explain supernova explosions over a wide mass range. We explore
possible issues that may affect the accuracy of numerical supernova
simulations, and review some of the ideas that have recently been
explored as avenues to robust explosions, including uncertainties in
the neutrino rates, rapid rotation, and an external forcing of
non-radial fluid motions by strong seed perturbations from convective
shell burning.  The ``perturbation-aided'' neutrino-driven mechanism
and the implications of recent 3D simulations of shell burning in
supernova progenitors are discussed in detail. The efficacy of the
perturbation-aided mechanism in some progenitors is illustrated by the first
successful multi-group neutrino hydrodynamics simulation of an $18
M_\odot$ progenitor with 3D initial conditions. We conclude with a few
speculations about the potential impact of 3D effects on the structure
of massive stars through convective boundary mixing.
\end{abstract}
\maketitle%
\section{INTRODUCTION}
\label{sec:intro}
The explosions of massive stars as \emph{core-collapse supernovae} (CCSNe)
constitute one of the most outstanding problems in modern astrophysics.
This is in no small measure due to the critical role of supernova
explosions in the history of the Universe. Core-collapse
supernovae figure prominently in the chemical evolution
of galaxies as the dominant producers, e.g., of
elements between oxygen and the iron group \citep{arnett_96,woosley_02},
and supernova feedback is a key ingredient in the modern
theory of star formation \citep{krumholz_14}.
The properties of neutron stars and stellar-mass black holes
(masses, spins, kicks; \citealt{oezel_10,oezel_12,kiziltan_13,antoniadis_16,arzoumanian_02,hobbs_05}) cannot be understood without
addressing the origin of these compact objects in stellar explosions.

Why (some) massive stars explode is, however, a daunting problem in
its own right regardless of the wider implications of supernova
explosions: The connection of supernovae of massive stars with the
gravitational collapse to a neutron star has been postulated more than
eighty years ago \citep{baade_34b}, and the best-explored mechanism
for powering the explosion, the neutrino-driven mechanism,
has gone through several stages of ``moulting'' in the fifty years
after its conception by \citet{colgate_66}. Yet the problem of the
supernova explosion mechanism still awaits a definitive solution. The
rugged path towards an understanding of the explosion mechanism merely
reflects that core-collapse supernovae are the epitome of a
``multi-physics'' problem that combines aspects of stellar structure
and evolution, nuclear and neutrino physics, fluid dynamics,
kinetic theory, and general relativity. We cannot recapitulate the
history of the field here and instead refer the reader to the
classical and modern reviews of
\citet{bethe_90}, \citet{arnett_96},
\citet{mezzacappa_05}, \citet{kotake_06}, \citet{janka_07}, \citet{burrows_07}, \citet{janka_12}, and \citet{burrows_13} as starting
points.  

The longevity of the supernova problem should not be misinterpreted:
Despite the occasional detour, supernova theory has made steady
progress, particularly so during the last few years, which have seen
the emergence of mature -- and increasingly successful --
multi-dimensional first-principle simulations of the collapse
and explosion of massive stars as well as conceptual advances
in our understanding of the neutrino-driven explosion mechanism and its
interplay with multi-dimensional hydrodynamic instabilities.

\subsection{The Neutrino-driven Explosion Mechanism
in its Modern Flavour} Before we review these recent advances, it is
apposite to briefly recapitulate the basic idea of the neutrino-driven
supernova mechanism in its modern guise. Stars with zero-age main
sequence masses above $\gtrsim 8 M_\odot$ and with a helium core mass
$\lesssim 65 M_\odot$ (the lower limit for non-pulsational
pair-instability supernovae; \citealp{heger_02,heger_03})
develop iron cores that eventually become subject to gravitational
instability and undergo collapse on a free-fall time-scale.  For
low-mass supernova progenitors with highly degenerate iron cores,
collapse is triggered by the reduction of the electron degeneracy
pressure due to electron captures; for more massive stars with higher
core entropy and a strong contribution of radiation pressure,
photo-disintegration of heavy nuclei also contributes
to gravitational instability.  Aside
from these ``iron core supernovae'', there may
also be a route towards core collapse from super-AGB stars with
O-Ne-Mg cores
\citep{nomoto_84,nomoto_87,poelarends_08,jones_13,jones_14,doherty_15},
where rapid core contraction is triggered by electron captures on
$^{20}\mathrm{Ne}$ and $^{24}\mathrm{Mg}$;\footnote{Whether the core
  continues to collapse to a neutron star depends critically on the
  details of the subsequent initiation and propagation of the oxygen
  deflagration during the incipient collapse
  \citep{isern_91,canal_92,timmes_92,schwab_15,jones_16}.} hence
this sub-class is designated as ``electron-capture supernovae'' (ECSNe).

According to modern shell-model calculations
  \citep{langanke_00,langanke_03}, the electron capture rate on heavy
  nuclei remains high even during the advanced stages of collapse
  \citep{langanke_03} when the composition of the core is dominated by
  increasingly neutron-rich and massive nuclei.  Further
deleptonisation during collapse thus reduces the lepton fraction
$Y_\mathrm{lep}$ to about $0.3$ according to modern simulations
\citep{marek_05,sullivan_16} until neutrino trapping occurs at a
density of $\mathord{\sim}10^{12} \, \mathrm{g} \, \mathrm{cm}^{-3}$.
As a result, the homologously collapsing inner core shrinks
\citep{yahil_83}, and the shock forms at a small enclosed mass of
$\mathord{\sim} 0.5 M_\odot$ \citep{langanke_03,hix_03,marek_05} after
the core reaches supranuclear densities and rebounds (``bounces'').
Due to
photodisintegration of heavy nuclei in the infalling shells into free
nucleons as well as rapid deleptonisation in the post-shock region
once the shock breaks out of the neutrinosphere,
the shock stalls a few milliseconds after bounce, i.e.\
it turns into an accretion shock with negative radial velocity downstream
of the shock.
Aided by a
continuous reduction of the mass accretion rate onto the young
proto-neutron star, the stalled accretion shock still propagates
outward for $\mathord{\sim} 70 \, \mathrm{ms}$, however, and reaches a typical
peak radius of $\mathord{\sim} 150 \, \mathrm{km}$ before it starts to
recede again.

The point of maximum shock expansion is roughly coincident with
several other important changes in the post-shock region: Photons and
electron-positron pairs become the dominant source of pressure in the
immediate post-shock region, deleptonisation behind the shock occurs
more gradually, and the electron neutrino and antineutrino
luminosities become similar. Most notably, a region of net neutrino
heating (``gain region'') emerges behind the shock. 
In the ``delayed
neutrino-driven mechanism'' as conceived by
\citet{bethe_85} and \citet{wilson_85}, the neutrino heating eventually leads to a
sufficient increase of the post-shock pressure to ``revive'' the shock
and make it re-expand, although the post-shock velocity initially
remains negative. Since shock expansion increases the mass of
the dissociated material exposed to strong neutrino heating, this is
thought to be a self-sustaining runaway process that eventually
pumps sufficient energy into the post-shock region to allow for
the development of positive post-shock velocities and, further
down the road, the expulsion of the stellar envelope.

Modern simulations of core-collapse supernovae that include
energy-dependent neutrino transport, state-of-the art microphysics,
and (to various degrees) general relativistic effects have
demonstrated that the neutrino-driven mechanism is not viable in
spherical symmetry
\citep{rampp_00,rampp_02,liebendoerfer_00,liebendoerfer_04,liebendoerfer_05,sumiyoshi_05,buras_06a,buras_06b,mueller_10,fischer_10,lentz_12a,lentz_12b},
except for supernova progenitors of the lowest masses
\citep{kitaura_06,janka_08,burrows_07b,fischer_10}, which will be
discussed in Section~\ref{sec:ecsn}.

In its modern guise, the paradigm of neutrino-driven explosions
therefore relies on the joint action of neutrino heating and various
hydrodynamic instabilities to achieve shock revival. As demonstrated
by the first generation of multi-dimensional supernova models in the
1990s \citep{herant_94,burrows_95,janka_95,janka_96}, the gain region
is subject to convective instability due to the negative entropy
gradient established by neutrino heating. Convection can be suppressed
if the accreted material is quickly advected from the shock to the
gain radius \citep{foglizzo_06}.  Under these conditions, the
standing accretion shock instability (SASI;
\citealp{blondin_03,blondin_06,foglizzo_07,laming_07,yamasaki_07,fernandez_09a,fernandez_09b}) can
still grow, which is mediated by an advective-acoustic cycle
\citep{foglizzo_02,foglizzo_07,guilet_12} and manifests itself in the
form of large-scale sloshing and spiral motions of the shock. The precise mechanism
whereby these instabilities aid shock revival requires careful
discussion (see Section~\ref{sec:3deffects}), but their net effect can be
quantified using the concept of the ``critical luminosity''
\citep{burrows_93} for the transition from a steady-state accretion
flow to runaway shock expansion: In effect, convection and/or the SASI
reduce the critical luminosity in multi-D by $20 \ldots 30 \%$
\citep{murphy_08b,nordhaus_10,hanke_12,fernandez_15} compared to the
case of spherical symmetry (1D).

\subsection{Current Questions and Structure of This Review}
We cannot hope to comprehensively review all aspects of the
core-collapse supernova explosion problem, even if we limit ourselves
to the neutrino-driven paradigm. Instead we shall focus
on the following topics that immediately connect to the above overview
of the neutrino-driven mechanism:

\begin{itemize}
\item The neutrino-driven explosion mechanism demonstrably works at
  the low-mass end of supernova progenitors. In
  Section~\ref{sec:ecsn}, we shall discuss the specific explosion
  dynamics in the region around the mass limit for iron core
  formation, i.e.\ for ECSN progenitors and structurally similar iron
  core progenitors. We shall also consider the nucleosynthesis in
  these explosions; since they are robust, occur early after bounce,
  and can easily be simulated until the explosion energy has
  saturated, explosions of ECSN and ECSN-like progenitors currently
  offer the best opportunity to study core-collapse supernova
  nucleosynthesis based on first-principle explosion models.
\item For more massive progenitors, it has yet to be demonstrated
that the neutrino-driven mechanism can produce robust explosions in 3D
with explosion properties (e.g.\ explosion energy, nickel mass, remnant mass)
that are compatible with observations. In Section~\ref{sec:3d}, we shall
review the current status of 3D supernova simulations, highlighting
the successes and problems of the current generation of models and detailing
the recent progress towards a quantitative understanding of the
interplay of neutrino heating and multi-dimensional fluid flow.
\item In the wake of a rapid expansion of the field of core-collapse
supernova modelling, a wide variety of methods
  have been employed to investigate the supernova problem with a
  continuum from a rigorous first-principle approach to parameterised
  models of limited applicability that are only suitable for attacking
  well-circumscribed problems. In Section~\ref{sec:methods}, we present
  an overview of the
  different numerical approaches to simulations of neutrino-driven
  explosions and provide some guidance for assessing and comparing
simulation results.
\item The problem of shock revival by the neutrino-driven mechanism has
not been conclusively solved. In Section~\ref{sec:prog}, we shall
review one of the promising ideas that could help explain
supernova explosions over a wide range of progenitors, viz.\
the suggestion that shock revival may be facilitated by
strong seed perturbations from prior convective
shell burning in the infalling O or Si shells
\citep{arnett_11,couch_13,mueller_15a,couch_15,mueller_16b}; and we
shall also discuss some other perspectives opened up
by current and future three-dimensional simulations of late
burning stages in supernova progenitors.
\end{itemize}

Potential observational probes for multi-dimensional fluid
flow in the supernova core during the first $\mathord{\sim} 1 \, \mathrm{s}$
exist in the form of the neutrino and gravitational wave signals,
but we shall not touch these in any depth and instead
point the reader to topical reviews
(\citealt{ott_08b,kotake_13} for gravitational wave emission;
\citealt{mirizzi_16} for the neutrino signal) as well as
some of the major publications of recent years
(gravitational waves: \citealt{mueller_13,yakunin_16,nakamura_16};
neutrinos: \citealt{tamborra_13,tamborra_14b,mueller_14})
Neither do we address alternative explosion scenarios here.
and refer the reader to \citet{janka_12} for a broader discussion
that covers, e.g.\, the magnetorotational mechanism as the
most likely explanation for hypernovae with explosion energies
of up to $\mathord{\sim} 10^{52} \, \mathrm{erg}$.

\section{THE LOW-MASS END ELECTRON-CAPTURE SUPERNOVAE AND THEIR COUSINS}
\label{sec:ecsn}
\begin{figure}
\begin{center}
\includegraphics[width=\linewidth]{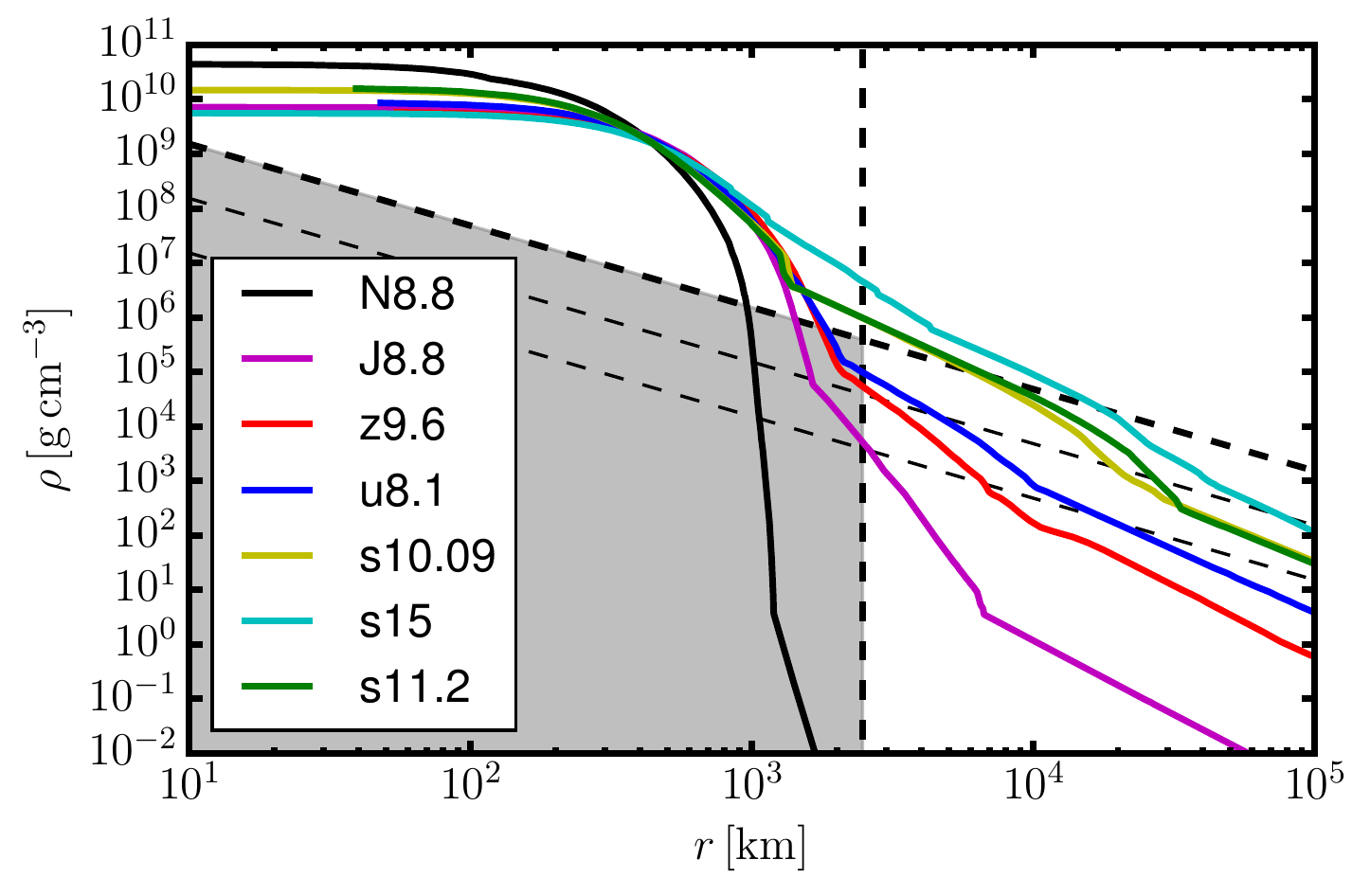}
\caption{Density profiles of several low-mass supernova
progenitors illustrating the conditions for ECSN-like explosions.
Profiles are shown for the $8.8 M_\odot$ ECSN-progenitor
of \citet{nomoto_84,nomoto_87} (N8.8, black),
the 8.8 $M_\odot$ ``failed massive star''
of \citet{jones_13} (J8.8, purple),
low-mass iron core progenitors (A.~Heger,
private communication)
of $9.6 M_\odot$ (z9.6, with Z=0, red) and
$8.1 M_\odot$ (u8.1, with $Z=10^{-4}$, blue), and
iron progenitors with $10.09 M_\odot$
and $15 M_\odot$ (s10.09 and s15, from \citealt{mueller_16a}, yellow
and cyan)
and $11.2 M_\odot$ (s11.2 from \citealt{woosley_02}, green).
The thick dashed vertical line roughly denotes
the location of the shell that reaches
the shock $0.5 \, \mathrm{s}$ after the
onset of collapse. Slanted dashed lines
roughly demarcate the regime where
the accretion rate onto the shock
reaches $0.05 M_\odot \, \mathrm{s}^{-1}$ (thick dashed line),
$5 \times 10^{-3} M_\odot \, \mathrm{s}^{-1}$ (thin),
and $5 \times 10^{-4} M_\odot \, \mathrm{s}^{-1}$ (thin)
(see Section~\ref{sec:ecsn_conditions} for
details and underlying assumptions). ECSN-like
explosion dynamics is expected if the density
profile intersects the grey region.
\label{fig:threshold}}
\end{center}
\end{figure}

Stars with zero-age main sequence (ZAMS) masses in the range
$\mathord{\sim} 8 \ldots 10 M_\odot$ exhibit structural peculiarities
during their evolution that considerably affect the supernova
explosion dynamics if they undergo core collapse.  The classical path
towards electron-capture supernovae (ECSNe;
\citealp{nomoto_84,nomoto_87}), where electron captures on ${}^{24}
\mathrm{Mg}$ and ${}^{20} \mathrm{Ne}$ in a degenerate O-Ne-Mg core of
$\mathord{\sim 1.37} M_\odot$ drive the core towards collapse, best
exemplifies these peculiarities: Only a small C/O layer is present on
top of the core, and the He layer has been effectively whittled down
by dredge-up. The consequence is an extremely steep density gradient
between the core and the high-entropy hydrogen envelope
(Figure~\ref{fig:threshold}). While this particular scenario is beset
with many uncertainties
\citep{siess_07,poelarends_08,jones_13,jones_14,jones_16,doherty_15,schwab_15,woosley_15},
recent studies of stellar evolution in the mass range around
$9M_\odot$ have demonstrated that there is a variety of paths towards
core-collapse that result in a similar progenitor structure
\citep{jones_13,woosley_15}, though there is some variation, e.g.\ in
the mass of the remaining He shell due to a different history of
dredge-up events. From the perspective of supernova explosion
dynamics, the crucial features in the mass range around $9M_\odot$ are
the small mass of the remaining C/O shell and the rapid drop of the
density outside the core; both are shared by ECSN
progenitors and the lowest iron-core progenitors.  This is
illustrated in Figure~\ref{fig:threshold} (see also Figure~7 in
\citealt{jones_13} and Figure~4 in \citealt{woosley_15}).

\subsection{Explosion Dynamics in ECSN-like Progenitors}
\subsubsection{Classical Electron-Capture Supernova Models}
\label{sec:classical}
The steep density gradient outside the core in ECSN-like progenitors
is immediately relevant for the dynamics of the ensuing supernova
because it implies a rapid decline of the mass accretion rate
$\dot{M}$ as the edge of the core reaches the stalled accretion
shock. A rapid drop in $\dot{M}$ implies a decreasing ram pressure
ahead of the shock and a continuously increasing shock radius (though
the shock remains a stationary accretion shock for at least
$\mathord{\sim} 50\, \mathrm{ms}$ after bounce and longer for some
ECSN-like progenitor models).  Under these conditions, neutrino
heating can easily pump sufficient energy into the gain region to make
the accreted material unbound and power runaway shock expansion.  As a
result, the neutrino-driven mechanism works for ECSN-like progenitors
even under the assumption of spherical symmetry. Using modern
multi-group neutrino transport, this was demonstrated by
\citet{kitaura_06} for the progenitor of \citet{nomoto_84,nomoto_87}
and confirmed in subsequent simulations by different groups
\citep{janka_08,burrows_07b,fischer_10}.  The explosions are
characterised by a small explosion energy of $\mathord{\sim} 10^{50}
\, \mathrm{erg}$ \citep{kitaura_06,janka_08} and a small nickel mass of a few $ 10^{-3}
M_\odot$ \citep{wanajo_09}.

Even though multi-dimensional effects are not crucial for shock
revival in these models, they are not completely negligible. Higher
entropies at the bottom of the gain layer lead to convective overturn
driven by Rayleigh-Taylor instability shortly after the explosion is
initiated \citep{wanajo_11}.  Simulations in axisymmetry (2D) showed
that this leads to a modest increase of the explosion energy in
\citet{janka_08}; an effect which is somewhat larger in more recent models
(von Groote et al., in preparation). The effect of
Rayleigh-Taylor overturn on the ejecta composition is, however, much
more prominent (see Section~\ref{sec:nucleosynthesis}).

\subsubsection{Conditions for ECSN-like Explosion Dynamics}
\label{sec:ecsn_conditions}
Not all of the newly available supernova progenitor models at the low-mass end
\citep{jones_13,jones_14,woosley_15} exhibit a similarly extreme
density profile as the model of \citet{nomoto_84,nomoto_87}; in some
of them the density gradient is considerably more shallow
(Figure~\ref{fig:threshold}).  This prompts the questions: How steep a
density gradient is required outside the core to obtain an explosion
that is triggered by a rapid drop of the accretion rate and works with
no or little help from multi-D effects? In reality, there will
obviously be a continuum between ECSN-like events and neutrino-driven
explosions of more massive stars, in which multi-D effects are crucial
for achieving shock revival.  Nonetheless, a rough distinction between
the two different regimes is still useful, and can be based on the
concept of the critical neutrino luminosity of \citet{burrows_93}.

\citet{burrows_93} showed that stationary accretion flow onto a
proto-neutron star in spherical symmetry is no longer possible if the
neutrino luminosity $L_\nu$ (which determines the amount of heating)
exceeds a critical value $L_\mathrm{crit}(\dot{M})$ that is well
approximated by a power law in $\dot{M}$ with a small exponent, or,
equivalently, if $\dot{M}$ drops below a threshold value for a given
luminosity. This concept has recently been generalised
\citep{janka_12,mueller_15a,summa_16,janka_16} to a critical relation for
the (electron-flavour) neutrino luminosity $L_\nu$ and neutrino mean
energy $E_\nu$ as a function of mass accretion rate $\dot{M}$ and
proto-neutron star mass $M$ as well as additional correction factors,
e.g., for shock expansion due to non-radial instabilities.

For low-mass progenitors with tenuous shells outside the core, $M$,
$L_\nu$, and $E_\nu$ do not depend
dramatically on the stellar structure outside the core during the early post-bounce: The
proto-neutron star mass is inevitably $M\approx 1.4 M_\odot$, and
since the neutrino emission is dominated by the diffusive neutrino
flux from the core, the neutrino emission properties are bound to be
similar to the progenitor of \citet{nomoto_84}, i.e.\ one has $L_\nu
\sim 5 \times 10^{52} \, \mathrm{erg} \, \mathrm{s}^{-1}$ and $E_\nu
\approx 11 \, \mathrm{MeV}$ \citep{huedepohl_10}, with a steady
decrease of the luminosity towards later times.  Using calibrated
relations for the ``heating
functional''\footnote{This compact designation
for $L_\nu E_\nu^2$ has been
suggested to me by H.-Th.~Janka.} $L_\nu E_\nu^2$ \citep{janka_16}, this translates into a critical mass accretion rate
of $\dot{M}_\mathrm{crit} \approx 0.07 M_\odot \, \mathrm{s}^{-1}$ for
ECSN-like progenitors. 

To obtain similarly rapid shock expansion as for the $8.8 M_\odot$
model of \citet{nomoto_84}, $\dot{M}$ must rapidly plummet \emph{well below}
this value. This can be translated into a condition for the density
profile outside the core using
analytic expressions for the infall time $t_\mathrm{infall}$
and accretion rate $\dot{M}$
for mass shell $m$,
which are roughly given by \citep{woosley_12,woosley_15b,mueller_16a},
\begin{equation}
t_\mathrm{infall}=\sqrt{\frac{\pi}{4 G \bar{\rho}}}
=\sqrt{\frac{\pi^2 r^3}{3 G m}},
\end{equation}
and
\begin{equation}
\label{eq:macc}
\dot{M}=\frac{2 m }{t_\mathrm{infall}} \frac{\rho}{\bar{\rho} - \rho},
\end{equation}
where $\bar{\rho}$ is the average density inside the mass shell.
For progenitors with little mass outside the core, we have
\begin{equation}
\dot{M}\approx\frac{2 m }{t_\mathrm{infall}} \frac{\rho}{\bar{\rho}}
=\frac{8\rho }{3}\sqrt{3 G m r^3}.
\end{equation}
Using $m=1.4 M_\odot$ and assuming that  $\dot{M}$ needs to
drop at least to $M_\mathrm{crit}=0.05 M_\odot \, \mathrm{s}^{-1}$ within $0.5 \, \mathrm{s}$
after the onset of collapse to obtain ECSN-like explosion dynamics,
one finds that the density needs to drop to
\begin{equation}
\rho \lesssim \frac{1}{8}\sqrt{\frac{3}{G m}}\dot{M}_\mathrm{crit}
r^{-3/2} 
\end{equation}
for a radius $r < 2230 \, \mathrm{km}$.

Figure~\ref{fig:threshold} illustrates that the density
gradient at the edge of the core can be far less
extreme than in the model of \citet{nomoto_84} to fulfil
this criterion. ECSN-like explosion dynamics
is expected alike for the modern
$8.8M_\odot$ ECSN progenitor of
\citet{jones_13} and low-mass iron cores
(A.~Heger, private communication)
of $8.1 M_\odot$ (with metallicity $Z=10^{-4}$)
and $9.6 M_\odot$ (Z=0), though the low-mass
iron core progenitors are a somewhat marginal case.

\subsubsection{Low-mass Iron Core Progenitors}
Simulations of these two low-mass iron progenitors with $8.1 M_\odot$
\citep{mueller_12b} and $9.6 M_\odot$ (\citealp{janka_12b,mueller_13}
in 2D; \citealp{melson_15a} in 3D) nonetheless demonstrated that the
structure of these stars is sufficiently extreme to produce
explosions reminiscent of ECSN models: Shock
revival sets in early around $100 \, \mathrm{ms}$ after bounce, aided
by the drop of the accretion rate associated with the infall of the
thin O and C/O shells, and the explosion energy remains small ($5
\times 10^{49} \ldots  10^{50} \, \mathrm{erg}$).

\begin{figure}
\begin{center}
\includegraphics[width=\linewidth]{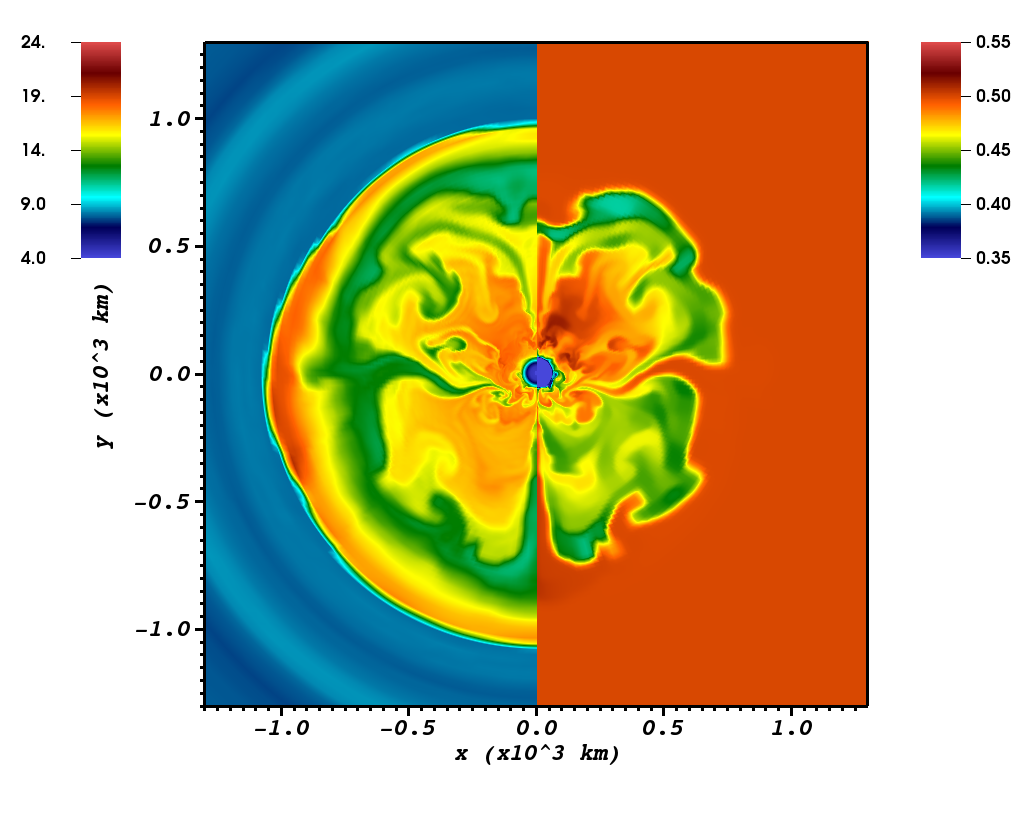}
\caption{Entropy $s$ (left half of plot) and electron fraction $Y_e$
  (right half) in the $9.6 M_\odot$ explosion model of
  \citet{janka_12b} and \citet{mueller_13} $280 \, \mathrm{ms}$ after
  bounce. Large convective plumes push neutron-rich material from
  close to the gain region out at high velocities.
\label{fig:z96_2d}}
\end{center}
\end{figure}

As shown by \citet{melson_15a}, there are important differences to
ECSNe, however: While shock revival also occurs
in spherical symmetry, multi-dimensional effects significantly alter
the explosion dynamics.  In 1D, the shock propagates very slowly
through the C/O shell after shock revival, and only accelerates
significantly after reaching the He shell. Without the additional
boost by convective overturn, the explosion energy is lower by a
factor of $\mathord{\sim} 5$ compared to the multi-D case.
Different from ECSNe, somewhat slower
shock expansion provides time for the small-scale convective plumes
to merge into large structures  as shown for
the $9.6 M_\odot$ model of \citet{janka_12b} in Figure~\ref{fig:z96_2d}.

Both for the $8.8 M_\odot$ model of \citet{wanajo_11} and the low-mass
iron-core explosion models, the dynamics of the Rayleigh-Taylor plumes
developing after shock revival is nonetheless quite similar. The
entropy of the rising plumes is roughly $\mathord{\sim} 15 \ldots 20
k_\mathrm{b}/\mathrm{nucleon}$ compared to $\mathord{\sim}10
k_\mathrm{b}/\mathrm{nucleon}$ in the ambient medium. For such an
entropy contrast, balance between buoyancy and drag forces applies a
limiting velocity of the order of the speed of sound. This limit
appears to be reached relatively quickly in the simulations. Apart
from the very early growth phase, the plume velocities should
therefore not depend strongly on the initial seed perturbations; they
are rather set by bulk parameters of the system, namely the post-shock
entropy at a few hundred kilometres and the entropy close to the gain
radius, which together determine the entropy contrast of the
plumes. This will become relevant later in our discussion of
the nucleosynthesis of ECSN-like explosions.

\begin{figure}
\begin{center}
\includegraphics[width=\linewidth]{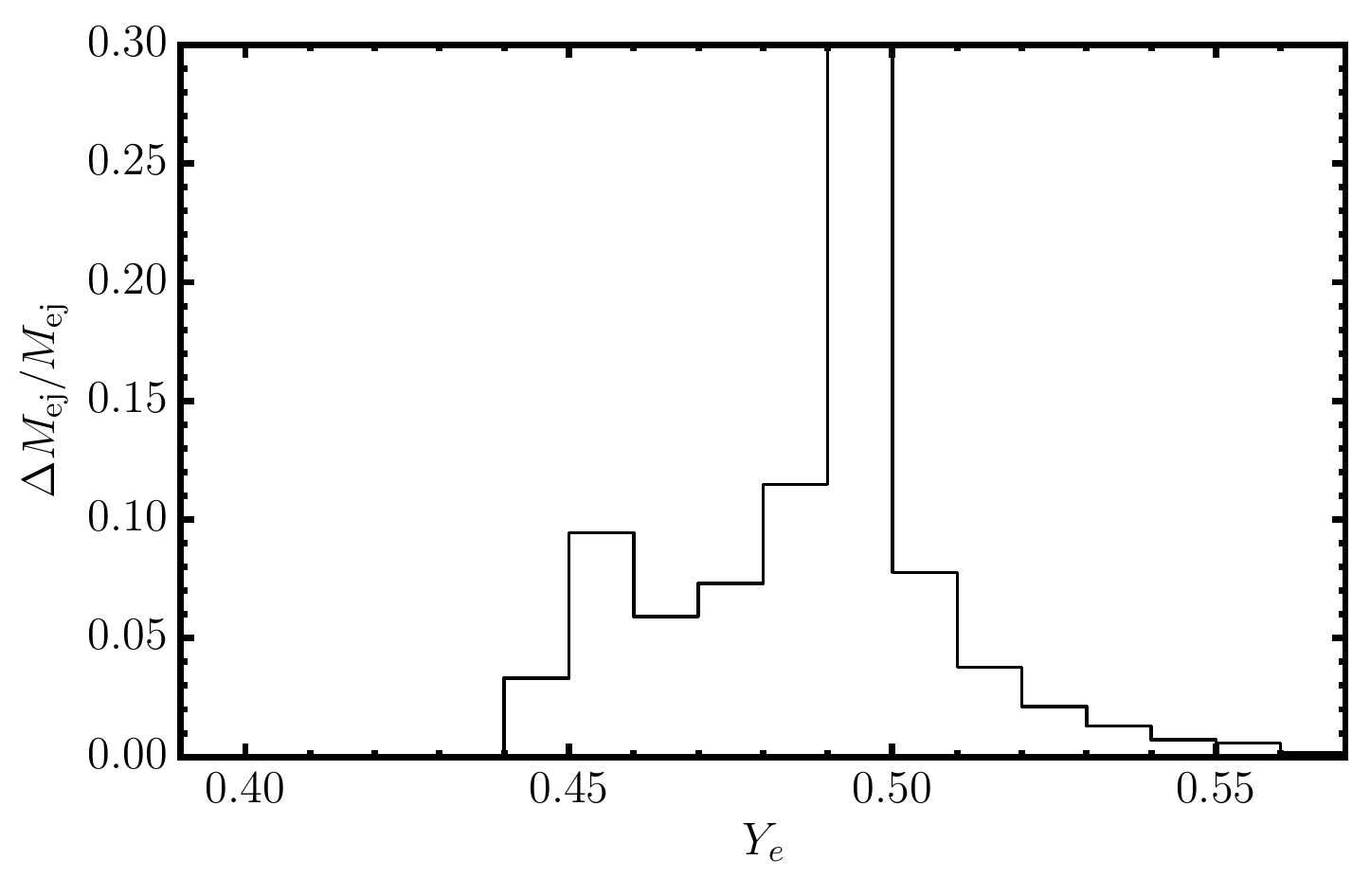}
\includegraphics[width=\linewidth]{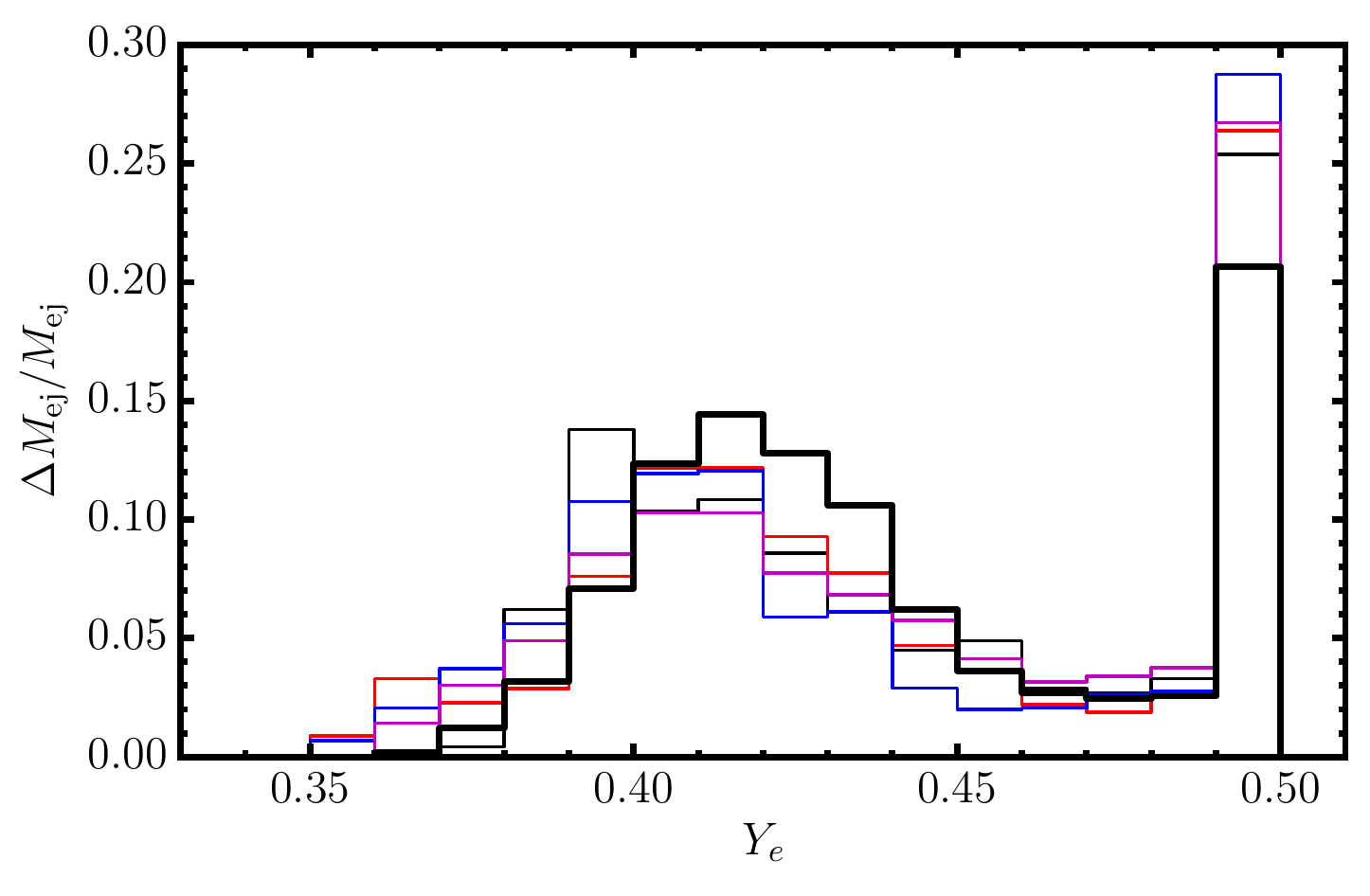}
\caption{Binned distribution of the electron fraction $Y_e$ in the
  early ejecta for different explosion models of a $9.6 M_\odot$ star
  $270 \, \mathrm{ms}$ after bounce.  The plots show the relative
  contribution $\Delta M_\mathrm{ej} /M_\mathrm{ej}$ to the total mass
  of (shocked) ejecta in bins with $\Delta Y_e=0.01$.  The upper panel
  shows the $Y_e$-distribution for the 2D model of \citet{janka_12b}
  computed using the \textsc{Vertex-CoCoNuT} code \citep{mueller_10}.
  The bottom panel illustrates the effect of stochastic variations and
  dimensionality using several 2D models (thin lines) and a 3D model
  computed with the \textsc{CoCoNuT-FMT} code \citet{mueller_15a} (thick lines).
  Note that the \emph{dispersion} in $Y_e$ in the early ejecta is
  similar for both codes, though the average $Y_e$ in the early ejecta
  is spuriously low when less accurate neutrino transport is used
  (\textsc{FMT} instead of \textsc{Vertex}).  The bottom panel is
  therefore only intended to show \emph{differential effects between
    different models}, and is not a prediction of the absolute value of
  $Y_e$. It suggests that i) stochastic variations
do not strongly affect the distribution of $Y_e$ in the ejecta,
and that ii) the resulting distribution of $Y_e$ in 2D and 3D is
relatively similar.
\label{fig:hists}}
\end{center}
\end{figure}

\subsection{Nucleosynthesis}
\label{sec:nucleosynthesis}
\subsubsection{1D Electron-Capture Supernovae Models -- Early Ejecta}
Nucleosynthesis calculations based on modern, spherically symmetric ECSN
models were first performed by \citet{hoffman_08} and
\citet{wanajo_09}. The results of these calculations appeared to point
to a severe conflict with observational constraints, showing a strong
overproduction of $N=50$ nuclei, in particular ${}^{90}\mathrm{Zr}$,
due to the ejection of slightly neutron-rich material (electron
fraction $Y_e\gtrsim 0.46$) with relatively low entropy ($s \approx 18
k_b/\mathrm{nucleon}$) immediately after shock revival.
\citet{hoffman_08} inferred that such nucleosynthesis yields
would only be compatible with chemogalactic evolution if ECSNe
were rare events occurring at a rate no larger than once per
3,000 years.

The low $Y_e$-values in the early ejecta stem from
the ejection of matter at relatively high velocities in
the wake of the fast-expanding shock. In slow outflows,
neutrino absorption
on neutrons and protons drives $Y_e$ to an equilibrium
value that is set by
the electron neutrino and antineutrino luminosities
$L_{\nu_e}$
and $L_{\bar{\nu}_e}$, the 
``effective'' mean energies\footnote{$\epsilon$
is given in terms
of the mean-square $\langle E^2\rangle$
and the mean energy $\langle E \rangle$,
as
$\epsilon=\langle E^2\rangle/
\langle E \rangle$. \citet{tamborra_12} can
be consulted for the ratio of the different
energy moments during various evolutionary phases.}
$\varepsilon_{\nu_e}$
and $\varepsilon_{{\bar\nu}_e}$,
and the proton-neutron mass difference
$\Delta =1.293 \, \mathrm{MeV}$
as follows \citep{qian_96},
\begin{equation}
Y_e \approx \left[1+\frac{L_{\bar{\nu}_e} (\varepsilon_{\bar{\nu}_e}-2 \Delta)}{L_{{\nu}_e} (\varepsilon_{{\nu}_e}+2 \Delta)}\right]^{-1}.
\end{equation}
For the relatively similar electron neutrino and antineutrino luminosities
and a small difference in the mean energies
of $2\ldots 3 \, \mathrm{MeV}$ in modern simulations, one
typically finds an asymptotic value of $Y_e>0.5$, i.e.\ \emph{proton-rich}
conditions. To obtain low $Y_e<0.5$ in the ejecta, neutrino
absorption reactions need to freeze out at a high density
(small radius) when
the equilibrium between the reactions $n(\nu_e,e^-)p$ and $p(\nu_e,e^+)n$
is still skewed towards low $Y_e$ due to electron captures
$p (e^-,\nu_e) n$ on protons. Neglecting
the difference between arithmetic, quadratic, and cubic
neutrino mean energies and assuming a roughly equal 
contribution of $n(\nu_e,e^-)p$ and $p(\bar{\nu}_e,e^+)n$
to the neutrino heating,
one can estimate that freeze-out roughly occurs when
(cp.\, Eq.~81 in \citealp{qian_96}),
\begin{equation}
\frac{v_r}{r} 
\approx \frac{2 m_N \dot{q}_\nu}{E_{\nu_e}+E_{\bar{\nu}_e}},
\end{equation}
where $m_N$ is the nucleon mass, 
$\dot{q}_\nu$ is the mass-specific neutrino heating
rate, $r$ is the radius and $v_r$ is the radial velocity.
Since $\dot{q}_\nu \propto r^{-2}$, freeze-out
will occur at smaller $r$, higher density,
and smaller $Y_e$ for higher ejection velocity.

\subsubsection{Multi-D Effects and the Composition
of the Early Ejecta} 
\label{sec:early_ejecta}
Since high ejection velocities translate into
lower $Y_e$, the Rayleigh-Taylor plumes in 2D simulations of ECSNe
(Figure~2 in \citealp{wanajo_11}) and explosions of low-mass iron
cores (Figure~\ref{fig:z96_2d}) contain material with even lower $Y_e$
than found in 1D ECSN models. Values of $Y_e$ as low as $0.404$ are found in
\citet{wanajo_11}.  

Surprisingly, \citet{wanajo_11} found that the neutron-rich plumes did
not aggravate the problematic overproduction of $N=50$ nuclei in their
2D ECSN model. This is due to the fact that the entropy in the
neutron-rich lumps is actually \emph{smaller} than in 1D\footnote{The
  dynamical reasons for this difference between 1D and multi-D models
  have yet to be investigated. Conceivably
shorter exposure to neutrino heating in 2D due to faster
expansion (which is responsible for the lower $Y_e$)
also decreases the final entropy of the ejecta.
} (but higher than in the ambient
medium), which changes the character of the
nucleosynthesis by reducing the $\alpha$-fraction at freeze-out from
nuclear statistical equilibrium (NSE). The result is an interesting
production of trans-iron elements between Zn and Zr for the progenitor
of \citet{nomoto_84,nomoto_87}; the production factors are consistent
with current rate estimates for ECSNe of about $4\%$ of all supernovae
\citep{poelarends_08}.  Subsequent studies showed that neutron-rich lumps in
the early ejecta of ECSNe could contribute a sizeable fraction to the
live $^{60}\mathrm{Fe}$ in the Galaxy \citep{wanajo_13b}, and might be
production sites for some other rare isotopes of obscure origin, such
as $^{48} \mathrm{Ca}$ \citep{wanajo_13a}.  Due to the similar
explosion dynamics, low-mass iron-core progenitors exhibit rather
similar nucleosynthesis (Wanajo et al., in preparation; Harris et al.,
in preparation). 
 The
results of these nucleosynthesis calculations tallies with the
observed abundance trends in metal-poor stars that suggest a separate
origin of elements like Sr, Y, and Zr from the heavy r-process
elements (light element primary process;
\citealp{travaglio_04,wanajo_06,qian_08,arcones_11,hansen_12,ting_12}).

Since $Y_e$ in the early ejecta of ECSNe and ECSN-like explosion is
sensitive to the neutrino luminosities and mean energies and to the
ejection velocity of the convective plumes (which may be different in
3D compared to 2D, or exhibit stochastic variations),
\citet{wanajo_11} also explored the effect of potential uncertainties
in the minimum $Y_e$ in the ejecta on the nucleosynthesis. They found
that a somewhat lower $Y_e$ of $\sim 0.3$ in the plumes might make
ECSNe a site for a ``weak r-process'' that could explain the enhanced
abundances of lighter r-process elements up to Ag and Pd in some
metal-poor halo stars \citep{wanajo_06,honda_06}.

Whether the neutron-rich conditions required for a weak r-process can
be achieved in ECSNe or low-mass iron-core supernovae remains to be
determined. Figure~\ref{fig:hists} provides a tentative glimpse on the
effects of stochasticity and dimensionality on the $Y_e$ in
neutron-rich plumes based on several 2D and 3D explosion models of a
$9.6 M_\odot$ low-mass iron core progenitor (A.~Heger,
  private communication) conducted using the \textsc{FMT} transport
scheme of \citet{mueller_15a}.\footnote{The \textsc{FMT} neutrino transport scheme cannot
  be relied upon for precise predictions of the value of $Y_e$, but should
be
  sufficiently accurate for exploring differential effects such as
  differences between plume expansion in 2D and 3D.}  Stochastic
variations in 2D models due to different (random) initial perturbations
shift the minimum $Y_e$ in the ejecta at most by
$0.02$. This is due to the fact that the Rayleigh-Taylor plumes
rapidly transition from the initial growth
phase to a stage where buoyancy and drag balance each other
and determine the velocity \citep{alon_95}. 3D effects do not
change the distribution of $Y_e$ tremendously either, at best they
tend to shift it to slightly higher values compared to 2D, which is
consistent with a somewhat stronger braking of expanding bubbles in 3D
as a result of the forward turbulent cascade \citep{melson_15a}.  It
thus appears unlikely that the dynamics of convective overturn is a
major source of uncertainty for the nucleosynthesis in ECSN-like
explosions, though confirmation with better neutrino transport
is still needed.

If these events are indeed sites of a weak r-process, the missing
ingredient is likely to be found elsewhere. Improvements in the
neutrino opacities, such as the proper inclusion of nucleon potentials
in the charged-current interaction rates
\citep{martinez_12,roberts_12c}, or flavour oscillations involving
sterile neutrinos \citep{wu_14} could lower $Y_e$ somewhat.
\citet{wu_14} found a significant reduction of $Y_e$ by up to $0.15$
in some of the ejecta, but these results may depend sensitively on the
assumption that collective flavour oscillations are still suppressed
during the phase in question. Moreover, \citet{wu_14} pointed out that
a reduction of $Y_e$ with the help of active-sterile flavour
conversion might require delicate fine-tuning to avoid shutting
off neutrino heating before the onset of the explosion due
to the disappearance of $\nu_e$'s (which could be fatal
to the explosion mechanism).

Moreover, whether ECSNe necessarily \emph{need} to co-produce Ag and
Pd with Sr, Y, and Zr is by no means clear. While observed abundance
trends may suggest such a co-production, the abundance patterns of
elements between Sr and Ag in metal-poor stars appear less robust
\citep{hansen_14}; and the failure of unaltered models to produce Ag
and Pd may not be indicative of a severe tension with observations.

\subsubsection{Other Nucleosynthesis Scenarios
for Electron-Capture Supernovae} There are at least two other
potentially interesting sites for nucleosynthesis in ECSN-like
supernovae.  For ``classical'' ECSN-progenitors with more extreme
density profiles, it has been proposed that the rapid acceleration of
the shock in the steep density gradient outside the core can lead to
sufficiently high post-shock entropies ($s \sim 100 \,
k_b/\mathrm{nucleon}$) and short expansion time-scales
($\tau_\mathrm{exp} \sim 10^{-4} \, \mathrm{s}$) to allow r-process
nucleosynthesis in the thin shells outside the core \citep{ning_07}.
This has not been borne out by numerical simulations, however
\citep{janka_08,hoffman_08}. When the requisite high entropy is
reached, the post-shock temperature has already dropped far too low to
dissociate nuclei, and the expansion time-scale does not become
sufficiently short for the scenario of \citet{ning_07} to work.
The proposed r-process in the rapidly expanding shocked shells
would require significantly different explosion dynamics, e.g.\ a
much higher explosion energy.

The neutrino-driven wind that is launched after accretion onto the
proto-neutron star has been completely subsided has long been
discussed as a potential site of r-process nucleosynthesis in
supernovae
\citep{woosley_94,takahashi_94,qian_96,cardall_97,thompson_01,arcones_07,arcones_13}.
ECSN-like explosions are in many respects the least favourable site
for an r-process in the neutrino driven wind since they produce
low-mass neutron stars, which implies low wind entropies and long
expansion time-scales \citep{qian_96}, i.e.\ conditions that are
detrimental to r-process nucleosynthesis. However, ECSNe are unique
inasmuch as the neutrino-driven wind can be calculated
self-consistently with Boltzmann neutrino transport
\citep{huedepohl_10,fischer_10} without the need to trigger an
explosion artificially. These simulations revealed a neutrino-driven
wind that is not only of moderate entropy ($s \lesssim 140
k_b/\mathrm{nucleon}$ even at late times), but also becomes
increasingly \emph{proton-rich} with time, in which case the $\nu
p$-process \citep{froehlich_06} could potentially operate.  The most
rigorous nucleosynthesis calculations for the neutrino-driven wind in
ECSNe so far \citep{pllumbi_15} are based on simulations that properly
account for nucleon interaction potentials in the neutrino opacities
\citep{martinez_12,roberts_12c} and have also explored the effects of
collective flavour oscillations, active-sterile flavour
conversion. \citet{pllumbi_15} suggest that wind nucleosynthesis in
ECSNe is rather mundane: Neither does the $\nu p$-process operate nor
can neutron-rich conditions be restored to obtain conditions even for
a weak r-process. Instead, they find that wind nucleosynthesis mainly
produces nuclei between Sc and Zn, but the production factors are low,
implying that the role of neutrino-driven winds in ECSNe is negligible
for this mass range for the purpose of chemogalactic evolution.

\subsection{Electron-Capture Supernovae -- Transients and Remnants}
Although the explosion mechanism of ECSNe is in many respects best
understood among all core-collapse supernova types from the viewpoint
of explosion mechanism, unambiguously identifying transients as ECSNe
has proved more difficult. It has long been proposed that SN~1054 was
an ECSN \citep{nomoto_82} based on the properties of its remnant, the
Crab nebula: The total mass of ejecta in the nebula is small
($\lesssim 5 M_\odot$; \citealp{davidson_85,macalpine_91,fesen_97}),
as is the oxygen abundance \citep{davidson_82,henry_82,henry_86},
which is in line with the thin O-rich shells in ECSN progenitors.
Moreover, the kinetic energy of the ejecta is only about $\lesssim
10^{50} \, \mathrm{erg}$ \citep{fesen_97,hester_08} as expected for an
ECSN-like event. Whether the Crab originates from a classical
ECSN or from something slightly different like a ``failed massive
star'' of \citet{jones_13} continues to be debated; 
\citet{macalpine_08} have argued, for example, against the former
interpretation based on a high abundance ratio of C vs.\ N and the
detection of some ashes of oxygen burning (S, Ar) in the nebula.

It has been recognised in recent years that the (reconstructed) light
curve of SN~1054 -- a type IIP supernova with a relatively bright
plateau -- is also compatible with the low explosion energy of
$\lesssim 10^{50} \, \mathrm{erg}$ predicted by recent numerical
simulations.  \citet{smith_13} interpreted the bright plateau,
which made SN~1054 visible by daytime for $\mathord{\sim} 3
\ \mathrm{weeks}$, as the result of interaction with circumstellar
medium (CSM). The scenario of \citet{smith_13} requires significant
mass loss ($0.1 M_\odot$ for about 30 years) shortly before the
supernova, which may be difficult to achieve, although
some channels towards ECSN-like explosions could involve
dramatic mass loss events \citep{woosley_15}. Subsequent
numerical calculations of ECSN light curves \citep{tominaga_13,moriya_14}
demonstrated, however, that less extreme assumptions for the mass loss
are required to explain the optical signal of SN~1054; indeed a very extended
hydrogen envelope may be sufficient to explain the bright plateau,
and CSM interaction with the progenitor wind may only be required
to prevent the SN from fading too rapidly.

Several other transients have also been interpreted as ECSNe,
e.g.\ faint type~IIP supernovae such as SN~2008S
\citep{botticella_09}.  \citet{smith_13} posits that ECSNe are
observed type IIn-P supernovae with circumstellar interaction like
SN~1994W with a bright plateau and a relatively sharp drop to a faint
nickel-powered tail, but again the required amount of
CSM is not easy to explain. All of these candidate
events share low kinetic energies and small nickel masses as a common
feature and are thus \emph{prima facie} compatible with ECSN-like
explosion dynamics.  Variations in the envelope structure of
ECSN-progenitors (e.g.\ envelope stripping in binaries) may account
for the very different optical signatures \citep{moriya_14}.

The peculiar nucleosynthesis in ECSNe-like explosions may also
leave observable fingerprints in the electromagnetic signatures.
The slightly neutron-rich character of the early ejecta
results in a strongly supersolar abundance ratio
of Ni to Fe after $\beta$-decays are completed  \citep{wanajo_11}.
Such high Ni/Fe ratios are seen in the nebular spectra of
some supernovae \citep{jerkstrand_15a,jerkstrand_15b}. ECSNe can only
explain some of these events, however; many of them
exhibit explosion energies and Nickel masses that are incompatible
with an ECSN.

\section{3D SUPERNOVA MODELS OF MASSIVE PROGENITORS}
\label{sec:3d}
In more massive progenitors with extended Si and O shells, the mass
accretion rate onto the shock does not drop as rapidly as in ECSN-like
explosions. Typically, one finds a relatively stable accretion rate of
a few $0.1 M_\odot \, \mathrm{s}^{-1}$ during the infall of the O
shell, which implies a high ram pressure ahead of the shock.  Under
these conditions, it is no longer trivial to demonstrate that neutrino
heating can pump a sufficient amount of energy into the post-shock
region to power runaway shock expansion.  1D simulations of the
post-bounce phase using Boltzmann solvers for the neutrino transport
convincingly demonstrated that neutrino-driven explosions cannot be
obtained under such conditions in spherical symmetry
\citep{liebendoerfer_00,rampp_00,thompson_00b}.  Much of the work of
recent years has therefore focused on better understanding and
accurately modelling how multi-dimensional effects in supernovae
facilitate neutrino-driven explosions -- an undertaking first begun in
the 1990s with axisymmetric (2D) simulations employing various
approximations for neutrino heating and cooling
\citep{herant_92,yamada_93,herant_94,burrows_95,janka_95,janka_96}.
2D simulations have by now matured to the point that multi-group
neutrino transport and the neutrino-matter interactions can be
modelled with the same rigour as in spherical symmetry
\citep{livne_04,buras_06a,mueller_10,bruenn_13,just_15,skinner_15}, or with,
still with acceptable accuracy for many purposes
(see Section~\ref{sec:methods} for a more careful discussion), by using some
approximations either in the transport treatment or the neutrino
microphysics \citep{suwa_10,mueller_15a,pan_16,oconnor_16,roberts_16}.

\subsection{Prelude -- First-principle 2D Models}
The current generation of 2D supernova simulations with multi-group
neutrino transport has gone a long way towards demonstrating that
neutrino heating can bring about explosion in conjunction with
convection or the SASI. Thanks to steadily growing computational
resources, the range of successful neutrino-driven explosion models
has grown from about a handful in mid-2012
\citep{buras_06b,marek_09,suwa_10,mueller_12a} to a huge sample of
explosion models with ZAMS masses between $10 M_\odot$ and $ 75
M_\odot$, different metallicities, and different choices for the
supranuclear equation of state
\citep{mueller_12b,janka_12b,suwa_13,bruenn_13,obergaulinger_14,nakamura_15,mueller_15b,bruenn_16,oconnor_16,summa_16,pan_16}.

Many of the findings from these simulations remain important and valid
after the advent of 3D modelling: The 2D models have established,
among other things, the existence of distinct SASI- and
convection-dominated regimes in the accretion phase, both of which can
lead to successful explosion \citep{mueller_12b} in agreement with
tunable, parameterised models \citep{scheck_08,fernandez_14}. They
have shown that ``softer'' nuclear equations of state that result
in more compact neutron stars are generally favourable
for shock revival \citep{janka_12,suwa_13,couch_12a}. The inclusion
of general relativistic effects, whether by means of
the conformally-flat approximation (CFC)
or, less rigorously, an effective pseudo-relativistic potential
for Newtonian hydrodynamics,
was found to have a similarly beneficial effect 
(CFC: \citealp{mueller_12a}; pseudo-Newtonian:
\citealp{oconnor_16}).
Moreover, there are signs that the 2D models of some groups
converge with each other; simulations of
four different stellar models ($12,15,20,25 M_\odot$) of
\citet{woosley_07} by
\citet{summa_16} and \citet{oconnor_16} have yielded
quantitatively similar results. 

Despite these successes, 2D models have, by and large, struggled to
reproduce the typical explosion properties of supernovae. They are
often characterised by a slow and unsteady growth of the explosion
energy after shock revival. Usually the growth of the explosion energy
cannot be followed beyond $2\ldots 4 \times 10^{50} \, \mathrm{erg}$
after simulating up to $\mathord{\sim} 1 \, \mathrm{s}$ of physical
time \citep{janka_12b,nakamura_15,oconnor_16}, i.e.\ below typical
observed values of $5\ldots 9 \times 10^{50} \, \mathrm{erg}$
\citep{kasen_09,pejcha_15c}. Only the models of \citet{bruenn_16}
reach significantly higher explosion energies.  While the explosion
energy often has not levelled out yet at the end of the simulations
and may still grow significantly for several seconds
\citep{mueller_15b}, its continuing growth comes at the expense of
long-lasting accretion onto the proto-neutron star. This may result
in inordinately high remnant masses. Thus, while
2D models appeared to have solved the problem of shock revival,
they faced an \emph{energy problem} instead.

\subsection{Status of 3D Core-Collapse Supernova Models}
Before 3D modelling began in earnest (leaving aside tentative sallies
into 3D by \citealt{fryer_02}), it was hoped that 3D effects might
facilitate shock revival even at earlier times than in 2D, and that
this might then also provide a solution to the energy problem, since
more energy can be pumped into the neutrino-heated ejecta at early
times when the mass in the gain region is larger.  These hopes were
already disappointed once several groups investigated the role of 3D
effects in the explosion mechanism using a simple ``light-bulb''
approach, where the neutrino luminosity and mean energy during the
accretion phase are prescribed and very simple approximations for the
neutrino heating and cooling terms are employed.  Although
\citet{nordhaus_10} initially claimed a significant reduction of the
critical neutrino luminosity for shock revival in 3D compared to 2D
based on such an approach, these results were affected by the gravity
treatment \citep{burrows_12} and have not been confirmed by subsequent
studies. Similar parameterised simulations have shown that the
critical luminosity in 3D is roughly equal to 2D
\citep{hanke_12,couch_12b,burrows_12,dolence_13} and about $20\%$ lower
than in 1D, though the results differ about the hierarchy
between 2D and 3D.

Subsequent supernova models based on multi-group neutrino transport
yielded even more unambiguous results: Shock revival in 3D was either
not achieved for progenitors that explode in 2D
\citep{hanke_13,tamborra_14a}, or was delayed significantly
\citep{takiwaki_14,melson_15b,lentz_15}. These first
disappointing results need to be interpreted carefully, however:
A detailed analysis of the heating conditions in the
 non-exploding 3D models of $11.2 M_\odot$, $20 M_\odot$,
and $27 M_\odot$ progenitors simulated by the Garching
supernova group revealed that these are very close to shock revival
\citep{hanke_13,hanke_phd,melson_15b}. Moreover, the 3D models
of the Garching group are characterised by
more optimistic heating conditions, larger average shock radii,
and higher kinetic energies in non-spherical motions compared
to 2D for extended periods of time; the same is
true for the delayed (compared to 2D) 3D explosion of \citet{lentz_15}
of  a $15M_\odot$ progenitor. It is merely when is comes
to sustaining shock expansion that the 3D models prove less
resilient than their 2D counterparts, which transition into
an explosive runaway more robustly.

The conclusion that 3D models are only slightly less prone to
explosion is reinforced by the emergence of the first successful
simulations of shock revival in progenitors with $20 M_\odot$
\citep{melson_15b} and $15 M_\odot$ \citep{lentz_15} using rigorous
multi-group neutrino transport and the best available neutrino
interaction rates. There is also a number of 3D explosion models
based on more simplified approaches to multi-group neutrino transport
\citep{takiwaki_12,takiwaki_14,mueller_15b,roberts_16}.

\subsection{How Do Multi-D Effects Facilitate Shock Revival?}
\label{sec:3deffects}
Despite these encouraging developments, several questions
now need to be addressed to make further progress:
What is the key to \emph{robust} 3D explosion models across
the entire progenitor mass range for which we observe
explosions (i.e.\ at least up to $15 \ldots 18 M_\odot$;
see \citealp{smartt_09a} and \citealp{smartt_15})? This question
is tightly connected to another, more fundamental one, namely: What
are the conditions for an explosive runaway, and 
how do multi-dimensional effects modify them? 

\subsubsection{Conditions for Runaway Shock Expansion}
Even without the complications of multi-D fluid flow, the physics of
shock revival is subtle.  In spherical symmetry, one can show that for
a given mass accretion rate $\dot{M}$, there is a maximum (``critical'')
electron-flavour luminosity $L_\nu$ at the neutrinosphere above which
stationary accretion flow onto the proto-neutron star is no longer
possible (\citealp{burrows_93}; cp.\ Section~\ref{sec:ecsn}).  This also
holds true if the contribution of the accretion luminosity due to
cooling outside the neutrinosphere is taken into account
\citep{pejcha_12}. The limit for the existence of stationary solutions
does not perfectly coincide with the onset of runaway shock
expansion, however. Using 1D light-bulb simulations (i.e.\ neglecting the
contribution of the accretion luminosity), \citet{fernandez_12} 
and \citet{gabay_15} showed
that the accretion flow becomes unstable to oscillatory and
non-oscillatory instability slightly below the limit of
\citet{burrows_93}. Moreover, it is unclear whether the negative
feedback of shock expansion on the accretion luminosity and hence on
the neutrino heating could push models into a limit cycle
(cp.\ Figure~28 of \citealp{buras_06a}) even above the threshold for
non-stationarity.  

Since an \emph{a priori} prediction of the critical luminosity, $L_\nu
(\dot{M})$ is not feasible, heuristic criteria have been developed
\citep{janka_98,janka_01b,thompson_00,thompson_05,buras_06b,murphy_08b,pejcha_12,fernandez_12,gabay_15,murphy_15}
to gauge the proximity of numerical supernova models to an explosive
runaway (rather than for pinpointing the formal onset of the runaway
after the fact, which is of less interest).  The most commonly used
criticality parameters are based on the ratio of two relevant
time-scales for the gain region
\citep{janka_98,janka_01b,thompson_00,thompson_05,buras_06b,murphy_08b}, namely the
advection or dwell time $\tau_\mathrm{adv}$ that accreted material
spends in the gain region, and the heating time-scale
$\tau_\mathrm{heat}$ over which neutrino energy deposition changes the
total or internal energy of the gain region appreciably.  If
$\tau_\mathrm{adv} > \tau_\mathrm{heat}$, neutrino heating can
equalise the net binding energy of the accreted material before it is
lost from the gain region, and one expects that the shock must expand
significantly due to the concomitant increase in pressure. Since
this expansion further increases $\tau_\mathrm{adv}$, an explosive runaway is
likely to ensue.

The time-scale criterion $\tau_\mathrm{adv}/\tau_\mathrm{heat}>1$ has
the virtue of being easy to evaluate since the two time-scales can be
defined in terms of global quantities such as the total energy
$E_\mathrm{tot,g}$ in the gain region, the volume-integrated neutrino
heating rate $\dot{Q}_\nu$, and the mass $M_\mathrm{g}$ in the gain
region (which can be used to define
$\tau_\mathrm{adv}=M_\mathrm{gain}/\dot{M}$ under steady-state
conditions). The significance of these global quantities for
the problem of shock revival is immediately intuitive, though
care must be taken to define the heating time-scale properly.
 \citet{thompson_00}, \citet{thompson_05}, \citet{murphy_08b}, and \citet{pejcha_12}
define  $\tau_\mathrm{heat}$ as the time-scale
for changes in the internal energy $E_\mathrm{int}$ in the
gain region,
\begin{equation}
\tau_\mathrm{heat}
=\frac{E_\mathrm{int}}{\dot{Q}_\nu},
\end{equation}
based on the premise that shock expansion is
regulated by the increase in pressure (and hence in internal energy).
This definition
yields unsatisfactory results, however. The criticality parameter
can be spuriously low at shock revival if this definition is used
($\tau_\mathrm{adv}/\tau_\mathrm{heat}<0.4$).

By defining $\tau_\mathrm{heat}$ in terms
of the total (internal+kinetic+potential) energy\footnote{Note
that rest-mass contributions to the internal energy are excluded
in this definition.}
of the gain region \citep{buras_06b},
\begin{equation}
\tau_\mathrm{heat}
=\frac{E_\mathrm{tot,g}}{\dot{Q}_\nu},
\end{equation}
the criterion $\tau_\mathrm{adv}/\tau_\mathrm{heat}>1$ becomes a very
accurate predictor for non-oscillatory instability
\citep{fernandez_12,gabay_15}. This indicates that the relevant energy
scale to which the quasi-hydrostatic stratification of the post-shock
region is the total energy (or perhaps the total or stagnation
enthalpy) of the gain region, and not the internal energy. This is
consistent with the observation that runaway shock expansion occurs
roughly once the total energy or the Bernoulli integral
\citep{fernandez_12,burrows_95} reach positive values somewhere 
(\emph{not} everywhere) in the
post-shock region, which is essentially what the time-scale criterion
estimates. What is crucial is that the density and pressure gradients
between the gain radius and the shock (and hence the shock position)
depends sensitively on the \emph{ratio} of enthalpy $h$ (or the
internal energy) and the gravitational potential, rather than on
enthalpy alone. Under the (justified) assumption
that quadratic terms in $v_r^2$ 
in the momentum and energy equation
are sufficiently small to be  neglected in the post-shock
region,
one can show (see Appendix~\ref{sec:app_gradient}) that the
logarithmic derivative of the density $\rho$ in the gain region is
constrained by
\begin{equation}
\frac{\pd \ln \rho}{\pd \ln r}
>-\frac{3GM}{rh},
\end{equation}
where $M$ is the proto-neutron star mass. Once $h>GM/r$ or even
$e_\mathrm{int}>GM/r$ (where $e_\mathrm{int}$ is the internal energy
per unit mass), significant shock expansion must ensue due to the
flattening of pressure and density gradients.

\citet{janka_12}, \citet{mueller_15a} and \citet{summa_16} have also pointed
out that the time-scale criterion can be converted into a scaling
law for the critical electron-flavour luminosity $L_\nu$
and mean energy $E_\nu$ in terms of the
proto-neutron star mass $M$, the accretion rate $\dot{M}$,
and the gain radius $r_\mathrm{g}$,
\begin{equation}
(L_\nu E_\nu^2)_\mathrm{crit} \propto (\dot{M} M)^{3/5}
r_\mathrm{g}^{-2/5}.
\end{equation}
The concept of the critical luminosity, the time-scale
criterion, and the condition of positive total energy
or a positive Bernoulli parameter at the gain radius are
thus intimately related and appear
virtually interchangeable considering that
they remain \emph{approximate criteria} for runaway shock expansion
anyway. This is also true for some other explosion criteria
that have been proposed, e.g.\ the antesonic condition of 
\citet{pejcha_12}, which states that the sound speed
$c_\mathrm{s}$ must exceed a certain fraction of the escape velocity $v_\mathrm{esc}$
for runaway shock expansion somewhere in the accretion flow,
\begin{equation}
c_\mathrm{s}^2>3/16 v_\mathrm{esc}^2.
\end{equation}
Approximating the equation of state as a radiation-dominated
gas with an adiabatic index $\gamma=4/3$ and a pressure
of $P=\rho e_\mathrm{int}  /3= \rho h/4$, one finds that
the antesonic condition roughly translates to
\begin{equation}
\frac{c_\mathrm{s}^2}{3/16 v_\mathrm{esc}^2}
=\frac{4/3 P/\rho}{3/8 GM/r}
=\frac{32 e_\mathrm{int}}{27 GM/r}
=\frac{8h}{9GM/r}>1,
\end{equation}
i.e.\ the internal energy and the enthalpy must be close to the
gravitational binding energy (even if the precise critical values for
$e_\mathrm{int}$ and $h$ may shift a bit for a realistic equation of
state).\footnote{This argument holds only for stationary 1D flow, however. In multi-D,
  the antesonic condition becomes sensitive to fluctuations in the
  sound speed, which limits its usefulness as diagnostic for the
  proximity to explosion. The fluctuations will be of order $\delta
  c_\mathrm{s}/c_\mathrm{s} \sim \delta \rho/\rho$, i.e.\ of the order of the square of
  the turbulent Mach number. This explains why high values of
  $c_\mathrm{s}^2/v_\mathrm{esc}^2$ are encountered in multi-D even in
  non-exploding models \citep{mueller_12a}. A similar problem
occurs if the shock starts to oscillate strongly in 1D close to
the runaway threshold.}

\subsubsection{Impact of Multi-D Effects on the Heating Conditions}
\label{sec:heat3d}
Why do multi-D effects bring models closer to shock revival, and how
is this reflected in the aforementioned explosion criteria?  Do these
explosion criteria even remain applicable in multi-D in the first
place?

The canonical interpretation has long been that the runaway
condition $\tau_\mathrm{adv}>\tau_\mathrm{heat}$ remains the decisive
criterion in multi-D, and that multi-D effects facilitate shock
revival mainly by increasing the advection time-scale
$\tau_\mathrm{adv}$ \citep{buras_06b,murphy_08b}. Especially close to
criticality, $\tau_\mathrm{heat}$ is also shortened due to feedback
processes -- better heating conditions imply that the net binding
energy in the gain region and hence $\tau_\mathrm{heat}$ must
decrease. 

While simulations clearly show increased advection time-scales in
multi-D compared to 1D \citep{buras_06b,murphy_08b,hanke_12} as a
result of larger shock radii, the underlying cause for larger
accretion shock radii in multi-D is more difficult to pinpoint.  Ever
since the first 2D simulations, both the transport of neutrino-heated
high-entropy material from the gain radius out to the shock
\citep{herant_94,janka_96} as well as the ``turbulent pressure'' of
convective bubbles colliding with the shock \citep{burrows_95} have
been invoked to explain larger shock radii in multi-D.  Both effects
are plausible since they change the components $P$ (thermal pressure)
and $\rho \mathbf{v} \otimes \mathbf{v}$ (where $\mathbf{v}$ is the
velocity) of the momentum stress tensor that must balance the ram
pressure upstream of the shock during stationary accretion.

That the turbulent pressure plays an important role follows already
from the high turbulent Mach number $\mathord{\sim 0.5}$ in the
post-shock region \citep{burrows_95,mueller_12b} before the onset of
shock revival, and has been demonstrated quantitatively by
\citet{murphy_12} and \citet{couch_14} using spherical Reynolds
decomposition to analyse parameterised 2D and 3D simulations.  Using a
simple estimate for the shock expansion due to turbulent pressure,
\citet{mueller_15a} were even able to derive the reduction of the
critical heating functional in multi-D compared to 1D
in terms of the average squared turbulent Mach number $\langle
\mathrm{Ma}^2\rangle$ in the gain region,
\begin{eqnarray}
\label{eq:lcrit_3d}
(L_\nu E_\nu^2)_\mathrm{crit,2D} 
&\approx&
(L_\nu E_\nu^2)_\mathrm{crit,1D}  \left(1+\frac{4\langle \mathrm{Ma}^2 \rangle}{3}\right)^{-3/5}\\
\nonumber &\propto& (\dot{M} M)^{3/5}
r_\mathrm{g}^{-2/5} 
 \left(1+\frac{4\langle \mathrm{Ma}^2 \rangle}{3}\right)^{-3/5},
\end{eqnarray}
and then obtained $(L_\nu E_\nu^2)_\mathrm{crit,2D} \approx 0.75
(L_\nu E_\nu^2)_\mathrm{crit,1D}$
in rough agreement with simulations using a model for the saturation of
non-radial fluid motions (see Section~\ref{sec:saturation}).

Nonetheless, there is likely no monocausal explanation for better
heating conditions in multi-D. \citet{yamasaki_06} found, for example,
that convective energy transport from the gain radius to the shock
also reduces the critical luminosity (although they somewhat
overestimated the effect by assuming constant entropy in the entire
gain region).  Convective energy transport reduces
the slope of the pressure gradient between the gain radius
(where the pressure is set by the neutrino luminosity and mean energy)
and the shock, and thus pushes the shock out by increasing the thermal
post-shock pressure. That this effect also plays a role
alongside the turbulent pressure can be substantiated by
an analysis of neutrino hydrodynamics simulations (Bollig et al.\, in preparation).

Only a detailed analysis of the properties of turbulence in the gain
region \citep{murphy_11} combined with a model for the interaction of
turbulence with a non-spherical accretion shock will reveal the
precise combination of multi-D effects that conspire to increase the
shock radius compared to 1D.  This is no prerequisite for
understanding the impact of multi-D effects on the runaway condition
as encapsulated by a phenomenological correction factor in
Equation~(\ref{eq:lcrit_3d}), since effects like turbulent energy
transport, turbulent bulk viscosity, etc.\ will also scale with the
square of the turbulent Mach number in the post-shock region just like
the turbulent pressure. They are
effectively lumped together in the correction factor $(1+4 /3 \langle
\mathrm{Ma}^2 \rangle)^{-3/5}$. \emph{The turbulent Mach number in
    the post-shock region is thus the crucial parameter for the
    reduction of the critical luminosity in multi-D},
although the coefficient of $\langle\mathrm{Ma}^2\rangle$ still needs to be calibrated
against multi-D simulations (and may be different in 2D and 3D).

This does \emph{not} imply, however, that the energetic requirements
for runaway shock expansion in multi-D are fundamentally different
from 1D: Runaway still occurs roughly once some material in the gain
region first acquires positive total (internal+kinetic+potential)
energy $e_\mathrm{tot}$; and the required energy input for this
ultimately stems from neutrino heating.\footnote{This is not at odds
  with the findings of \citet{murphy_08b} and \citet{couch_14}, who
  noticed that the neutrino heating rate in light-bulb and
  leakage-based multi-D simulations at runaway is \emph{smaller} than
  in 1D.  Due to a considerably different pressure and density
  stratification (cf.\ Figure~3 in \citealt{couch_14}, which shows a
  very steep pressure gradient behind the shock in the critical 1D
  model), the gain region needs to become much more massive in 1D than in multi-D
  before the runaway condition
  $\tau_\mathrm{adv}/\tau_\mathrm{heat}>1$ is met. Therefore
  \emph{both} the neutrino heating rate $\dot{Q}_\nu$ and the binding
  energy $E_\mathrm{tot}$ of the gain region are higher around shock
  revival in 1D (as both scale with $M_\mathrm{gain}$).}

\begin{figure}
\begin{center}
\includegraphics[width=\linewidth]{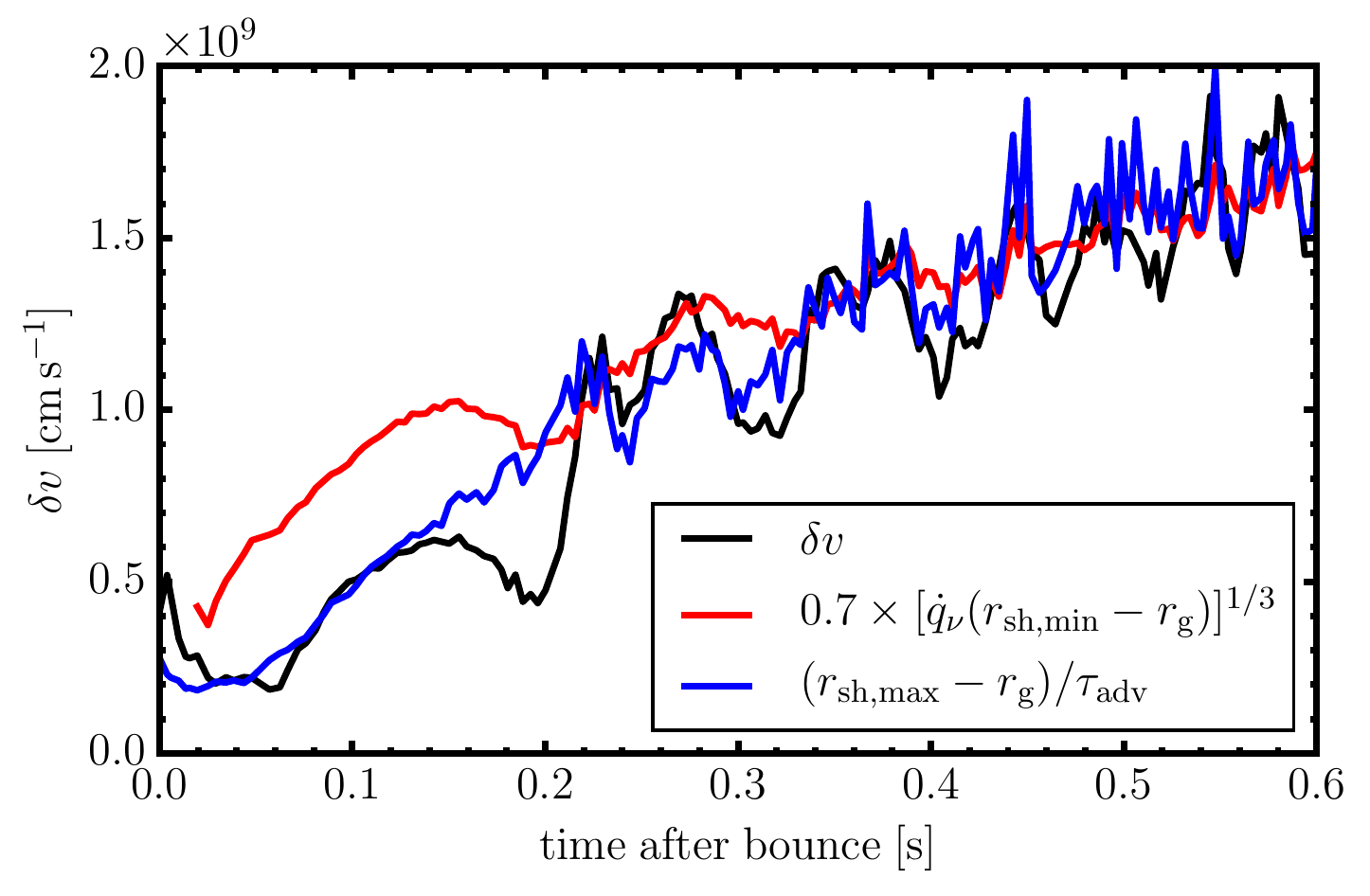}
\caption{Comparison of the
root-mean-square average $\delta v$ of non-radial velocity component
in the gain region (black) with two phenomenological models for the saturation of non-radial
instabilities in a SASI-dominated 3D model of an $18 M_\odot$ star
using the \textsc{CoCoNuT-FMT} code. The red curve shows an estimate based on
Equation~(\ref{eq:vturb}), which rests on 
the assumption of a balance between
buoyant driving and turbulent dissipation \citep{murphy_12,mueller_15a}.
The blue curve shows the prediction of
Equation~(\ref{eq:vsasi}), which assumes that saturation 
is regulated by a balance between the growth rate of the SASI
and parasitic Kelvin-Helmholtz instabilities \citep{guilet_10}.
Even though Equation~(\ref{eq:vsasi}) assumes a constant quality
factor $|\mathcal{Q}|$ to estimate the SASI growth rate,
it appears to provide a good estimate for the dynamics of
the model. Interestingly, the saturation models for
the SASI- and convection dominated regimes give similar results
during later phases even though the mechanism behind the driving
instability is completely different.
\label{fig:vturb3d}}
\end{center}
\end{figure}

\subsubsection{Saturation of Instabilities}
\label{sec:saturation}
What complicates the role of multi-D effects in the neutrino-driven
mechanism is that the turbulent Mach number in the gain region itself
depends on the heating conditions, which modify the growth rates and
saturation properties of convection and the SASI. Considerable
progress has been made in recent years in understanding this
feedback mechanism and the saturation properties of these two
instabilities.

The \emph{linear} phases of convection and the SASI
are now rather well understood. The growth rates for buoyancy-driven convective
instability are expected to be of order of the Brunt-V\"ais\"al\"a
frequency
$\omega_\mathrm{BV}$, which can be expressed in terms
of $P$, $\rho$, $c_\mathrm{s}$, and the local gravitational acceleration
$g$ as\footnote{Note that different sign conventions
for $\omega_\mathrm{BV}$ are used in the literature;
here $\omega_\mathrm{BV}^2>0$ corresponds to instability.}
\begin{equation}
\omega_\mathrm{BV}^2=g \left(\frac{1}{\rho } \frac{\pd \rho}{\pd r}-
\frac{1}{\rho c_\mathrm{s}^2}\frac{\pd P}{\pd r}\right),
\end{equation}
which becomes positive in the gain region due to neutrino heating.
A first-order estimate yields,
\begin{equation}
\omega_\mathrm{BV}^2 \sim 
\frac{G M \dot{Q}_\nu}{4\dot{M} r_\mathrm{g} ^2 c_\mathrm{s}^2 \left(r_\mathrm{sh}-r_\mathrm{g}\right)}
\sim \frac{3\dot{Q}_\nu}{4\dot{M} r_\mathrm{g} \left(r_\mathrm{sh}-r_\mathrm{g}\right)},
\end{equation}
using $c_\mathrm{s}^2 \approx GM/(3r_\mathrm{g})$ at the gain radius \citep[cp.\ ][]{mueller_15a}.
An important subtlety is that advection can stabilise the
flow so that
$\omega_\mathrm{BV}^2>0$ is no longer sufficient for instability
unless large seed perturbations in density are already present.
Instability instead depends on the more
restrictive criterion for
the parameter $\chi$ \citep{foglizzo_06},
\begin{equation}
\chi = \int\limits_{r_\mathrm{g}}^{r_\mathrm{sh}}\frac{\omega_\mathrm{BV}}{|v_r|}\ud r,
\end{equation}
with $\chi \gtrsim 3$ indicating convective instability.

The scaling of the linear growth rate  $\omega_\mathrm{SASI}$ of SASI modes is more complicated,
since it involves both the duration $\tau_\mathrm{cyc}$
of the underlying advective-acoustic cycle as well
as a quality factor $\mathcal{Q}$ for the conversion
of vorticity and entropy perturbations into
acoustic perturbation in the deceleration region
below the gain region and the reverse process
at the shock \citep{foglizzo_06,foglizzo_07},
\begin{equation}
\label{eq:om_sasi}
\omega_\mathrm{SASI}
\sim \frac{ \ln |\mathcal{Q}|} {\tau_\mathrm{cyc}}.
\end{equation}
For realistic models with strong SASI, one finds $\ln |\mathcal{Q}|
\sim 2$ \citep{scheck_08,mueller_12b}.  SASI growth appears to be
suppressed for $\chi \gtrsim 3$ probably because convection destroys
the coherence of the waves involved in the advective-acoustic cycle
\citep{guilet_10}.  Interestingly, the demarcation line $\chi=3$
between the SASI- and convection-dominated regimes is also valid in
the non-linear regime if $\chi$ is computed from the angle- and
time-averaged mean flow \citep{fernandez_12}; and both the SASI and
convection appear to drive $\chi$ close to this critical value
\citep{fernandez_12}.

Both in the SASI-dominated regime and the convection-dominated regime,
large growth rates are observed in simulations.  It only takes a few
tens of milliseconds until the instabilities reach their saturation
amplitudes. For this reason, the turbulent Mach number and the
beneficial effect of multi-D effects on the heating conditions are
typically more sensitive to the saturation mechanism than to initial
conditions, so that the onset of shock revival is only subject to
modest stochastic variations \citep{summa_16}.  Exceptions apply when
the heating conditions vary rapidly, e.g.\ due to the infall of a
shell interface or extreme variations in shock radius (as in the
light-bulb models of \citealt{cardall_15}), and the runaway condition
is only narrowly met or missed \citep{melson_15a,roberts_16}.

The saturation properties of convection were clarified by
\citet{murphy_12}, who determined that the volume-integrated
neutrino heating rate $\dot{Q}_\nu$ and the convective
luminosity  $L_\mathrm{conv}$ in the gain region roughly balance
each other. This can be understood as the result
of a self-adjustment process of the accretion flow,
whereby a marginally stable, quasi-stationary stratification with
$\chi \approx 3$ is established \citep{fernandez_12}.
\citet{mueller_15a} showed that this can be translated into a
scaling law that relates the average mass-specific neutrino heating
rate $\dot{q}_\nu$ in the gain region to
 the root mean square average
$\delta v$
of non-radial velocity fluctuations,
\begin{equation}
\label{eq:vturb}
\delta v \sim \left[ \dot{q}_\nu (r_\mathrm{sh}-r_\mathrm{g}) \right]^{1/3}.
\end{equation}

That a similar scaling should apply in the SASI-dominated regime is
not immediately intuitive. \citet{mueller_15a} in fact tested
Equation~(\ref{eq:vturb}) using a SASI-dominated 2D model and argued
that self-adjustment of the flow to $\chi \approx 3$ will result in
the same scaling law as for convection-dominated models. However, 
models suggest that a different mechanism may be at play in the
SASI-dominated regime. Simulations are at least equally
compatible with the mechanism proposed by
\citet{guilet_10}, who suggested that saturation of the SASI
is mediated by parasitic instabilities and occurs once the
growth rate of the parasite equals the growth rate of the SASI:
Assuming that the Kelvin-Helmholtz instability is the dominant
parasite, a simple order-of-magnitude estimate for saturation
can be obtained by equating $\omega_\mathrm{SASI}$
and the average shear rate,
\begin{equation}
\omega_\mathrm{SASI} 
\sim \frac{\delta v}{\Lambda}
\end{equation}
where $\Lambda$ is the effective width of the shear
layer. \citet{kazeroni_16} find that the Kelvin-Helmholtz
instability operates primarily in directions where
the shock radius is larger, which suggests
$\Lambda =r_\mathrm{sh,max}-r_\mathrm{g}$.
This results in a scaling law that
relates the velocity fluctuations
to the average radial velocity  $\langle v_r \rangle$ in the gain region,
\begin{equation}
\label{eq:vsasi}
\delta v \sim \omega_\mathrm{SASI} \Lambda
\sim 
\frac{\ln |\mathcal{Q}| (r_\mathrm{sh,max}-r_\mathrm{g})}{\tau_\mathrm{adv}}
\sim
\ln |\mathcal{Q}| \, |\langle v_r \rangle|,
\end{equation}
where we assumed $\tau_\mathrm{cyc} \approx \tau_\mathrm{adv}$.
The quality factor $\mathcal{Q}$ can in principle change significantly
with time and between different models.
Nonetheless, together with
the assumption of a roughly constant quality
factor, Equation~(\ref{eq:vsasi}) appears to capture
the dynamics of the SASI in 3D quite well for a simulation
of an $18 M_\odot$ progenitor with the \textsc{CoCoNuT-FMT}
code \citep{mueller_15a} as illustrated in Figure~\ref{fig:vturb3d}.

Equation~(\ref{eq:vturb}) for the convection-dominated
regime and Equation~(\ref{eq:vsasi}) apparently
predict turbulent Mach numbers in the same ballpark.
This can be understood by expressing
$\dot{q}_\nu$ in terms
of the accretion efficiency $\eta_\mathrm{acc}
=L_\nu/(GM \dot{M}/r_\mathrm{g})$
and the heating efficiency $\eta_\mathrm{heat}
=\dot{Q}_\nu/L_\nu$,
\begin{eqnarray}
\dot{q}_\nu &=&\frac{\dot{Q}_\nu}{M_\mathrm{g}}=
 \eta_\mathrm{heat} \eta_\mathrm{acc}
\frac{GM \dot{M}} {r_\mathrm{g} M_\mathrm{g}}=
 \eta_\mathrm{heat} \eta_\mathrm{acc}
\frac{GM} {r_\mathrm{g} \tau_\mathrm{adv}}
\\
\nonumber
&=&
\eta_\mathrm{heat} \eta_\mathrm{acc}
\frac{GM}{r_\mathrm{sh} \tau_\mathrm{adv}}\frac{r_\mathrm{sh}}{r_\mathrm{gain}}.
\end{eqnarray}
If we neglect the ratio $r_\mathrm{sh}/r_\mathrm{g}$
and approximate the average post-shock velocity
as $| \langle v_\mathrm{r} \rangle | \approx \beta^{-1} 
\sqrt{GM/r_\mathrm{sh}}$
(where $\beta$ is the compression ratio
in the shock), we obtain
\begin{equation}
\dot{q}_\nu \sim
\eta_\mathrm{heat} \eta_\mathrm{acc}
\frac{\beta^2 |\langle v_r \rangle|^2 }{\tau_\mathrm{adv}},
\end{equation}
and hence
\begin{equation}
\delta v \sim (\eta_\mathrm{heat} \eta_\mathrm{acc} \beta^2)^{1/3}
|\langle v_r \rangle|.
\end{equation}
For plausible values (e.g.\ $\eta_\mathrm{heat} = 0.05$
$\eta_\mathrm{acc} =2$,
$\beta=10$), one finds $\delta v \sim 2 |\langle v_r \rangle|$,
i.e.\ the turbulent Mach number at saturation is of the
same order of magnitude in the convection- and SASI-dominated
regimes (where at least $\ln |\mathcal{Q}| \mathord{\sim} 2$ can
be reached).

Equations~(\ref{eq:vturb}) and (\ref{eq:vsasi}) remain
order-of-magnitude estimates; and either of the instabilities may be
more efficient at pumping energy into non-radial turbulent motions in
the gain region, as suggested by the light-bulb models of \citet{fernandez_15}
and \citet{cardall_15}. These authors find that the SASI can lower the
critical luminosity in 3D considerably further than
convection. \citet{fernandez_10} attributes this to the emergence of
the spiral mode of the SASI \citep{blondin_07,fernandez_10} in 3D,
which can store more non-radial kinetic energy than the SASI sloshing
mode in 2D, but this has yet to be borne
out by self-consistent neutrino hydrodynamics simulations
(see Section~\ref{sec:outlook} for further discussion).

\subsubsection{Why Do Models Explode More Easily in 2D Than in 3D?}
How can one explain the different behaviour of 2D and 3D models in the
light of our current understanding of the interplay between neutrino
heating, convection, and the SASI? It seems fair to say that
we can presently only offer a heuristic interpretation for
the more pessimistic evolution of 3D models.

The most glaring difference between 2D and 3D models (especially
in the convection-dominated regime) prior to shock revival lies in the typical
scale of the turbulent structures, which are smaller
in 3D \citep{hanke_12,couch_12b,couch_14}, whereas
the inverse turbulent cascade in 2D \citep{kraichnan_67} artificially
channels turbulent kinetic energy to large scales. 
This implies that the effective dissipation length (or also
the effective mixing length for energy transport) are smaller
in 3D, so that smaller dimensionless coefficients
$C$ appear in relations like Equation~(\ref{eq:vturb}),
\begin{equation}
\label{eq:vturb1}
\delta v =C \left[ \dot{q}_\nu (r_\mathrm{sh}-r_\mathrm{g})
  \right]^{1/3},
\end{equation}
and the turbulent Mach number will be smaller for a given neutrino
heating rate. Indeed, for the $18 M_\odot$ model shown in
Figure~(\ref{fig:vturb3d}), we find
\begin{equation}
\label{eq:vturb3}
\delta v =0.7 \left[ \dot{q}_\nu (r_\mathrm{sh}-r_\mathrm{g})
  \right]^{1/3}
\end{equation}
in 3D rather than what
\citet{mueller_15a} inferred from 2D models (admittedly
using a different progenitor),
\begin{equation}
\label{eq:vturb2}
\delta v =\left[ \dot{q}_\nu (r_\mathrm{sh}-r_\mathrm{g})
  \right]^{1/3}.
\end{equation}
Following the arguments of \citet{mueller_15a} to infer 
the correction factor 
$\left(1+\frac{4\langle \mathrm{Ma}^2 \rangle}{3}\right)^{-3/5}$
for multi-D effects in Equation~(\ref{eq:lcrit_3d}), one
would then expect a
\emph{considerably} larger critical luminosity in 3D, i.e.\
$(L_\nu E_\nu^2)_\mathrm{crit,3D}
\approx 0.85  (L_\nu E_\nu^2)_\mathrm{crit,1D}$
instead of
$(L_\nu E_\nu^2)_\mathrm{crit,2D}
\approx 0.75  (L_\nu E_\nu^2)_\mathrm{crit,1D}$ in 2D.

Such a large difference in the critical luminosity does not tally with
the findings of light-bulb models that show that the critical
luminosities in 2D and 3D are still very close to each other. This
already indicates that more subtle effects may be at play in 3D
that almost compensate the stronger effective dissipation of turbulent
motions. The fact that simulations typically show transient phases of
stronger shock expansion and more optimistic heating conditions in 3D
than in 2D \citep{hanke_12,melson_15b} also points in this direction.

Furthermore, light-bulb models \citep{handy_14} and multi-group
neutrino hydrodynamics simulations \citep{melson_15a,mueller_15b} have
demonstrated that favourable 3D effects come into play \emph{after
  shock revival}. These works showed that 3D effects can lead
to a faster, more robust growth of the explosion energy provided
that shock revival can be achieved in the first place.

The favourable 3D effects that are responsible for this may already
counterbalance the adverse effect of stronger dissipation in the
pre-explosion phase to some extent: Energy leakage from the gain
region by the excitation of g-modes is suppressed in 3D because the
forward turbulent cascade \citep{melson_15a} and (at high Mach number)
the more efficient growth of the Kelvin-Helmholtz instability
\citep{mueller_15b} brake the downflows before they penetrate the
convectively stable cooling layer. Moreover, the non-linear growth of
the Rayleigh-Taylor instability is faster for three-dimensional
plume-like structures than for 2D structures with planar
\citep{yabe_91,hecht_95,marinak_95} or toroidal geometry (as in the
context of Rayleigh-Taylor mixing in the stellar envelope during the
explosion phase; \citealp{kane_00,hammer_10}), which might explain why
3D models initially respond more strongly to sudden drops in the
accretion rate at shell interfaces and exhibit better heating
conditions than their 2D counterparts for brief periods.
Finally, the difference in the effective dissipation
length in 3D and 2D that is reflected by
Equations~(\ref{eq:vturb3}) and (\ref{eq:vturb2}) may not
be universal and depend, e.g., on the heating conditions
or the $\chi$-parameter;  the results of \citet{fernandez_15}
in fact demonstrate that under appropriate circumstances
more energy can be stored in non-radial motions in 3D than
in 2D in the SASI-dominated regime.

\subsection{Outlook: Classical Ideas for More Robust Explosions}
\label{sec:outlook}
The existence of several competing -- favourable and unfavourable --
effects in 3D first-principle models does not change the
fundamental fact that they remain more reluctant to explode than their
2D counterparts.  This suggests that some important physical
ingredient are still lacking in current simulations.  Several avenues
towards more robust explosion models have recently been explored.
Some of the proposed solutions have a longer pedigree and revisit
ideas (rapid rotation in supernova cores, enhanced neutrino
luminosities) that have been investigated on and off in supernova theory
already before the advent of 3D simulations. The more
``radical'' solution of invoking strong seed perturbations from
convective shell burning to boost non-radial instabilities in
the post-shock region will be discussed separately in
Section~\ref{sec:prog}.

\subsubsection{Rotation and Beyond}
\citet{nakamura_14} and \citet{janka_16} pointed out that rapid
progenitor rotation can facilitate explosions in 3D.  \citet{janka_16}
ascribed this partly to the reduction of the pre-shock infall velocity
due to centrifugal forces, which decreases the ram pressure ahead of
the shock. Even more importantly, rotational support also decreases
the net binding energy $|e_\mathrm{tot}|$ per unit mass in the gain
region in their models. They derived an analytic correction factor for
the critical luminosity in terms of the average specific angular
momentum $j$ in the infalling shells,
\begin{equation}
\label{eq:lcrit_rot}
(L_\nu E_\nu^2)_\mathrm{crit,rot} \approx
  (L_\nu E_\nu^2)_\mathrm{crit} \times \left(1-\frac{j^2}{2 G M
    r_\mathrm{sh}}\right)^{3/5}.
\end{equation}
Assuming rapid rotation with $j \gg 10^{16} \, \mathrm{cm}^2\,
\mathrm{s}^{-1}$, one can obtain a significant reduction of the
critical luminosity by several $10 \%$ as \citet{janka_16} tested in
a simulation with a modified rotation profile.\footnote{One should
  bear in mind, though, that rotation also decreases the neutrino
  luminosity and mean neutrino energy because it leads to larger
  neutron star radii \citep{marek_09}.} 
For very rapid rotation, other explosion mechanisms also become
feasible, such as the magnetorotational mechanism
\citep{akiyama_03,burrows_07b,winteler_12,moesta_14},
or explosions driven by the low-$T/W$ spiral instability
\citep{takiwaki_16}.

However, current stellar evolution models do not predict the required
 rapid rotation rates for these scenarios for the generic progenitors
 of type~IIP supernovae. The typical specific angular momentum at a
 mass coordinate of $m=1.5 M_\odot$ is only of the order of $j \sim
 10^{15} \, \mathrm{cm}^2 \, \mathrm{s}^{-1}$ in models
 \citep{heger_05} that include angular momentum transport by magnetic
 fields generated by the Tayler-Spruit dynamo \citep{spruit_02}, and
 asteroseismic measurements of core rotation in evolved low-mass stars
 suggest that the spin-down of the cores may be even more efficient 
\citep{cantiello_14}.
 For such slow rotation, centrifugal forces are negligible;
 Equation~(\ref{eq:lcrit_rot}) suggests a change of the critical
 luminosity on the per-mil level.  Neither is rotation expected to
 affect the character of neutrino-driven convection appreciably because
the angular velocity $\Omega$ in the gain region is too small.
The Rossby number is well above unity,
\begin{equation}
\mathrm{Ro}\sim \frac{|v_r|}{ (r_\mathrm{sh}-r_\mathrm{g}) \Omega}
\sim \frac{r_\mathrm{s}^2}{\tau_\mathrm{adv} j}
\sim 10,
\end{equation}
assuming typical values of $\tau_\mathrm{adv} \sim 10 \, \mathrm{ms}$
and $r_\mathrm{sh} \sim 100\, \mathrm{km}$.

Magnetic field amplification by a small-scale dynamo
or the SASI \citep{endeve_10,endeve_12} could also help to facilitate
shock revival with magnetic fields acting as a \emph{subsidiary}
to neutrino heating but without directly powering the
explosion as in the magnetorototational mechanism.
The 2D simulations of \citet{obergaulinger_14} demonstrated that magnetic
fields can help organise the flow into large-scale
modes and thereby allow earlier explosions, though the
required initial field strengths for this are higher
($\mathord{\sim} 10^{12} \, \mathrm{G}$) than
the typical values predicted by stellar evolution models.

\subsubsection{Higher Neutrino Luminosities and Mean Energies?}
Another possible solution for the problem of missing or delayed
explosions in 3D lies in increasing the electron flavour luminosity
and mean energy. This is intuitive from Equation~(\ref{eq:lcrit_3d}),
where a mere change of $\mathord{\sim}5\%$ in both $L_\nu$ and
$E_\nu$ results in a net effect of $16\%$, which is almost on par with
multi-D effects.

The neutrino luminosity is directly sensitive to the neutrino
opacities, which necessitates precision modelling in order to capture
shock propagation and heating correctly
(\citealp{lentz_12a,lentz_12b,mueller_12a}; see also
Section~\ref{sec:methods}), as well as to other physical ingredients
of the core-collapse supernova problem that influence the contraction
of the proto-neutron star, such as general relativity and the nuclear
equation of state
\citep{janka_12,mueller_12a,couch_12a,suwa_13,oconnor_16}.  Often such
changes to the neutrino emission come with counterbalancing side
effects (\emph{Mazurek's law}); e.g.\ stronger neutron star
contraction will result in higher neutrino luminosities and mean
energies, but will also result in a more tightly bound gain region,
which necessitates stronger heating to achieve shock revival.

That the lingering uncertainties in the microphysics may nonetheless
hold the key to more robust explosions has long been recognised in the
case of the equation of state.  \citet{melson_15b} pointed out that
missing physics in our treatment of neutrino-matter interactions may
equally well be an important part of the solution of the problem shock
revival. Exploring corrections to neutral-current scattering cross
section due to the ``strangeness'' of the nucleon, they found that
changes in the neutrino cross section on the level of a few ten
percent were sufficient to tilt the balance in favour of explosion for
a $20 M_\odot$ progenitor. While \citet{melson_15b} deliberately
assumed a larger value for the contribution of strange quarks to the
axial form factor of the nucleon than currently measured
\citep{airapetian_07}, the deeper significance of their result is that
Mazurek's law can sometimes be circumvented so that modest changes in
the neutrino opacities still exert an appreciable effect on supernova
dynamics. A re-investigation of the rates currently employed in the
best supernova models for the (more uncertain) neutrino interaction
processes that depend strongly on in-medium effects (charged-current
absorption/emission, neutral current scattering, Bremsstrahlung;
\citealp{burrows_98,burrows_99,reddy_99,hannestad_98}) may thus be
worthwhile (see \citealp{bartl_14,rrapaj_15,shen_14} for some recent
efforts).

\section{ASSESSMENT OF SIMULATION METHODOLOGY}
\label{sec:methods}

Considering what has been pointed out in Section~\ref{sec:3d} -- the
crucial role of hydrodynamic instabilities and the delicate
sensitivity of shock revival to the neutrino luminosities and mean
energies -- it is natural to ask: What are the requirements for
modelling the interplay of the different ingredients of the
neutrino-driven mechanism accurately? This question is even more
pertinent considering that the enormous expansion of the field during
the recent years has sometimes produced contradictory results, debates
about the relative importance of physical effects, and controversies
about the appropriateness of certain simulation methodologies.

Ultimately, only the continuous evolution of the simulation codes, the
inclusion of similar physics by different groups, and carefully
designed cross-comparisons will eventually produce a ``concordance
model'' of the neutrino-driven mechanism and confirm that simulation
results are robust against uncertainties.  For 1D neutrino
hydrodynamics simulations, this has largely been achieved in the wake
of the pioneering comparison paper of \citet{liebendoerfer_05}, which
has served as reference for subsequent method papers and sensitivity
studies in 1D
\citep{mueller_10,lentz_12a,lentz_12b,oconnor_15,just_15,summa_16}.
Similar results of the Garching-QUB collaboration \citep{summa_16} and
\citet{oconnor_16} with multi-group neutrino transport indicate a
trend to a similar convergence in 2D, and more detailed comparisons
are underway (see, e.g.,
\url{https://www.authorea.com/users/1943/articles/97450/_show_article}
for efforts coordinated by E.~O'Connor).  Along the road to
convergence, it appears useful to provide a preliminary review of some
issues concerning the accuracy and reliability of supernova
simulations.

\subsection{Hydrodynamics}
Recently, the discussion of the fidelity of the simulations has
strongly focused on the the hydrodynamic side of the problem.
As detailed in Section~\ref{sec:3d}, multi-D effects play a crucial role in the explosion mechanism,
and  are regulated by a balance of driving
(by neutrino heating through buoyancy, or by an inherent instability
of the flow like the SASI) and dissipation. 

\subsubsection{Turbulence in Supernova Simulations}
This balance needs to be modelled with sufficient physical and
numerical accuracy. On the numerical side, the challenge consists in
the turbulent high-Reynolds number flow, and the question arises to
what extent simulations with relatively coarse resolution can capture
this turbulent flow accurately.  Various authors
\citep{handy_14,abdikamalov_15,radice_15,roberts_16} have stressed that the regime
of fully developed turbulence cannot be reached with the limited
resolution affordable to cover the gain region ($\mathord{\sim} 100$
zones, or even less) in typical models, and
\citet{handy_14} thus prefer to speak of
``perturbed laminar flow'' in simulations. Attempts to quantify the effective
Reynolds number of the flow using velocity structure functions
and spectral properties of the post-shock turbulence
\citep{handy_14,abdikamalov_15,radice_15} put it at a few
hundred at best, and sometimes even below $100$. 

This is in line with rule-of-thumb estimates based on the numerical
diffusivity for the highest-wavenumber (odd-even) modes in
Godunov-based schemes as used in many supernova codes.  This
diffusivity can be calculated analytically (Appendix~D of
\citealt{mueller_phd}; see also \citealt{arnett_16} for a simpler
estimate).  For Riemann solvers that take all the wave families into
account (e.g.\,
\citealp{colella_85,toro_94,mignone_05_a,marquina_96}), the numerical
kinematic viscosity $\nu_\mathrm{num}$ in the subsonic regime is
roughly given in terms of the typical velocity jump per cell $\delta v_\mathrm{gs}$ and the cell width $\delta l$ as
$\nu_\mathrm{num} \sim \delta l \, \delta v_\mathrm{gs}$.  Relating
$\delta v_\mathrm{gs}$ to the turbulent velocity $v$ and scale $l$ of
the largest eddy as $\delta v_\mathrm{gs} \sim v (\delta l/l)^{1/3}$
(i.e.\ assuming Kolmogorov scaling)
yields a numerical Reynolds number of
\begin{equation}
\mathrm{Re} = \frac{v l}{\nu_\mathrm{num}} \sim
\left(\frac{l}{\delta l}\right)^{4/3}=N^{4/3},
\end{equation}
where $N$ is the number of zones covering the largest eddy scale.  For
more diffusive solvers like HLLE \citep{einfeldt_88}, one obtains
$\nu_\mathrm{num} \sim \delta l \, c_\mathrm{s} \sim \delta l \, v\,
\mathrm{Ma}^{-1}$ instead and 
\begin{equation}
\mathrm{Re} \sim (l/\delta l)
\mathrm{Ma} \sim N \, \mathrm{Ma},
\end{equation}
i.e.\ such solvers are strongly inferior for subsonic flow with
low Mach number $\mathrm{Ma}$.

Such coarse estimates are to be taken with caution, however.  The
numerical dissipation is non-linear and self-regulated as typical of
implicit large-eddy simulations (ILES, \citealp{boris_92,grinstein_07}). In
fact, the estimates already demonstrate that simply comparing the
resolution in codes with different solvers and grid geometries can be
misleading. Codes with three-wave solvers like \textsc{Vertex-Prometheus}
\citep{rampp_02,buras_06a} and \textsc{CoCoNuT-FMT}
\citep{mueller_15a} of the MPA-QUB collaboration, \textsc{Flash}
\citep{fryxell_00} as used in \citet{couch_12a} and subsequent work by
S.~Couch and E.~O'Connor, and the VH-1 hydro module \citep{blondin_91}
in the \textsc{Chimera} code of the Oak~Ridge-Florida
Atlantic-NC State collaboration, have less
stringent resolution requirements than HLLE-based codes
\citep{ott_12,kuroda_12}. The reconstruction method, special
tweaks for hydrostatic equilibrium (or an the lack of such a
treatment), as well as the grid geometry and grid-induced
perturbations \citep{janka_16,roberts_16} also affect the behaviour
and resolution-dependence of the simulated turbulence.

\subsubsection{Resolution Requirements -- A Critical Assessment}
Regardless of the employed numerical schemes, the fact remains that
the achievable numerical Reynolds number in supernova simulations is
limited, and that the regime of fully developed turbulence
($\mathrm{Re} \gg 1000$) will not be achieved in the near future, as
it would require $\gtrsim 512$ radial zones \emph{in the gain region
  alone}. The question for supernova models, however, is not whether all
the facets of turbulence in inviscid flow can be reproduced, but
whether the flow properties that matter for the neutrino-driven
mechanism are computed with sufficient accuracy. 
In fact, one cannot
even hope that simply cranking up the numerical resolution with ILES
methods would give the correct solution: In reality, non-ideal
effects such as neutrino viscosity and drag
\citep{van_den_horn_84,burrows_88,jedamzik_98,guilet_15} come into
play, and deviations of the turbulent Prandtl number from unity as well as MHD
effects like a small-scale dynamo (see Section~\ref{sec:outlook})
can complicate the picture even for non-rotating, weakly magnetised
supernova cores. These effects  will likely not grossly alter
the dynamics of convection and the SASI, but the physical
reality may be slightly different from the limit of infinite resolution
if these effects are not accounted for and inviscid flow is assumed instead.

At the end of the day, these additional complications and the finite
resolution probably have a limited effect on supernova dynamics, since
they only affect \emph{a correction term} to the critical luminosity
such as $(1+4 /3 \langle \mathrm{Ma}^2 \rangle)^{-3/5}$ in
Equation~(\ref{eq:lcrit_3d}) through
the effective dissipation length that determines
the non-dimensional coefficient in Equation~(\ref{eq:vturb}).
If we repeat the analytic estimate for
$L_\mathrm{crit}$ of \citet{mueller_15a}, but assume stronger
dissipation and decrease their critical Mach number at shock revival
$\mathrm{Ma}_\mathrm{crit}^2=0.4649$ by $10\%$, then
Equation~(\ref{eq:lcrit_3d}) suggests an increase of the critical
luminosity from $74.9\%$ of the 1D value to of $76.6\%$ of the 1D value,
which is a minute change.  Modelling turbulent dissipation within $10 \%$
uncertainty thus seems wholly sufficient given that one can hardly
hope to achieve $1\%$ accuracy for the neutrino luminosities and mean
energies. 

The turbulent dissipation does not change without bounds with
increasing resolution, but eventually reaches an asymptotic limit
at high Reynolds numbers. Although most supernova
simulation may not fully reach this asymptotic regime, they do not
fall far short of it:
The works of \citet{handy_14} and \citet{radice_15,radice_16} suggest
that this level of accuracy in the turbulent dissipation can be
reached even with moderate resolution ($<100$ grid points per
direction, $\sim 2^\circ$ resolution in angle in spherical
polar coordinates) in the gain region with higher-order reconstruction methods
and accurate Riemann solvers. 
Problems due to stringent resolution
requirements may still lurk elsewhere, though, e.g.\ concerning SASI
growth rates as already pointed out ten years ago by \citet{sato_08}.
Resolution studies and cross-comparisons thus remain useful, though
cross-comparisons are of course hampered by the different physical
assumptions used in different codes and the feedback processes in the
supernova core. For this reason a direct comparison of, e.g.,
turbulent kinetic energies and Mach numbers between different models
is not necessarily meaningful. The dimensionless coefficients
governing the dynamics of non-radial instabilities such the
proportionality constant $\eta_\mathrm{conv}=v_\mathrm{turb}/[\dot{q_\nu}
(r_\mathrm{sh}-r_\mathrm{g})]$ in Equation~(\ref{eq:vturb}) or the
quality factor $\mathcal{Q}$ in Equation~(\ref{eq:om_sasi}) may be
more useful metrics of comparison.

\subsection{Neutrino Transport}
The requirements on the treatment of neutrino heating and cooling are
highly problem-dependent. The \emph{physical principles} behind
convection and the SASI can be studied with simple heating and cooling
functions in a light-bulb approach, and such an approach is indeed
often advantageous as it removes some of the feedback processes that
complicate the analysis of full-scale supernova simulations.  To model
the fate and explosion properties of concrete progenitors in a
predictive manner, some form of neutrino transport is required, and
depending on the targeted level of accuracy, the requirements become
more stringent; e.g.\ higher standards apply when it comes to
predicting supernova nucleosynthesis. There is no perfect method for
neutrino transport in supernovae as yet. Efforts toward a solution of
the full 6-dimensional Boltzmann equation are underway
\citep[e.g.\ ][]{cardall_13,peres_13,radice_13,nagakura_14}, but not yet ripe for
real supernova simulations.

Neutrino transport algorithms (beyond fully parameterised light-bulb
models) currently in use for 1D and multi-D models include:
\begin{itemize}
\item leakage schemes
as, e.g., in \citet{oconnor_10}, \citet{oconnor_11}, \citet{ott_13} and \citet{couch_14b}
\item  the isotropic
diffusion source approximation  (IDSA) of \citet{liebendoerfer_09},
\item one-moment closure schemes employing prescribed flux factors
\citep{scheck_06}, flux-limited diffusion as
in the \textsc{Vulcan} code \citep{livne_04,walder_05}, 
the \textsc{Chimera} code \citep{bruenn_85,bruenn_13}
and the \textsc{Castro} code \citep{zhang_13,dolence_15}, or
a dynamic closure as in the \textsc{CoCoNuT-FMT} code,
\item two-moment
methods employing algebraic closures
in 1D \citep{oconnor_15} and multi-D
\citep{obergaulinger_11,kuroda_12,just_15,skinner_15,oconnor_16,roberts_16,kuroda_16}
or variable Eddington factors from a model Boltzmann equation 
\citep{burrows_00,rampp_02,buras_06a,mueller_10},
\item discrete ordinate methods for the Boltzmann equation,
mostly in 1D \citep{mezzacappa_93,yamada_99,liebendoerfer_04}
or, at the expense of other simplifications, in multi-D (\citealp{livne_04,ott_08_a,nagakura_16};
only for static configurations: \citealp{sumiyoshi_15}).
\end{itemize}

This list should not be taken as a hierarchy of accuracy; it mere
reflects crudely the rigour in treating \emph{one aspect} of the
neutrino transport problem, i.e.\, the angle-dependence of the
radiation field in phase space. When assessing neutrino transport
methodologies, there are other, equally important factors that need to
be taken into account when comparing different modelling approaches.

Most importantly, the sophistication of the microphysics varies
drastically. On the level of one-moment and two-moment closure models, it
is rather the neutrino microphysics that decides about the
quantitative accuracy. The 3D models of the MPA-QUB group
\citep{melson_15a,melson_15b,janka_16} and the \textsc{Chimera} team
\citep{lentz_15} currently represent the state-of-the-art in this
respect; though other codes \citep{oconnor_15,just_15,skinner_15,kuroda_16} come
close.

Often, the neutrino physics is simplified considerably, however.  Some
simulations disregard heavy flavour neutrinos altogether
\citep[e.g][]{suwa_10,takiwaki_12}, or only treat them by means of a
leakage scheme \citep{takiwaki_14,pan_16}. This affects the
contraction of the proto-neutron star and thus indirectly alters the
emission of electron flavour neutrinos and the effective inner
boundary for the gain region as well.

Among multi-D codes, energy transfer due to inelastic
  neutrino-electron scattering (NES) is routinely taken into account
  only in the \textsc{Vertex} code
  \citep{rampp_02,buras_06a,mueller_10} of the MPA-QUB collaboration,
  the \textsc{Alcar} code \citep{just_15}, the \textsc{Chimera} code
  of the \textsc{Chimera} team \citep{bruenn_85,bruenn_13}, and the
  \textsc{Fornax} code of the Princeton group \citep{skinner_15}.
  Without NES \citep{bruenn_85} and modern electron capture rates
  \citep{langanke_03}, the core mass at bounce is larger and the shock
  propagates faster at early times \citep{lentz_12a,lentz_12b}. In
  multi-D, this can lead to unduly strong prompt convection. Because
  of this problem, a closer look at the bounce dynamics is in order
  whenever explosions occur suspiciously early ($< 100 \ \,
  \mathrm{ms}$ after bounce).  Parameterising deleptonisation during
  collapse \citep{liebendoerfer_05_b} provides a workaround to some
  extent.

The recoil energy transfer in neutrino-nucleon scattering
effectively reshuffles heavy flavour neutrino luminosity
to electron flavour luminosity in the cooling region
\citep{mueller_12a} and hence critically influences the
heating conditions in the gain region. Among multi-D codes, only
\textsc{Vertex} and \textsc{Chimera} currently take this into account,
and the code \textsc{CoCoNuT-FMT} \citep{mueller_15a} uses
an effective absorption opacity for heavy flavour neutrinos to mimic this
phenomenon.

\textsc{Vertex} and \textsc{Chimera} are also the only multi-D codes
to include the effect of nucleon-nucleon correlations
\citep{burrows_98,burrows_99,reddy_99} on absorption and scattering
opacities. Nucleon correlations have a huge impact during the cooling
phase, which they shorten by a factor of several \citep{huedepohl_10}.
Their role during the first second after bounce is not well
explored. Considering that the explosion energetics are determined on
a time-scale of seconds \citep{mueller_15b,bruenn_16}, it is plausible
that the increased diffusion luminosity from the neutron star due
to in-medium corrections to the opacities may influence the explosion
energy to some extent.

Gray schemes \citep{fryer_02,scheck_06,kuroda_12} cannot model
neutrino heating and cooling accurately; an energy-dependent treatment
is needed because of the emerging neutrino spectra are
highly non-thermal with  a pinched high-energy tail \citep{janka_89b,keil_03}.

Some multi-D codes use the ray-by-ray-plus approximation \citep{buras_06a},
which exaggerates angular variations in the radiation field, and has been claimed to
lead to spuriously early explosions in some cases in conjunction
with artificially strong sloshing motions in 2D \citep{skinner_15}. Whether
this is a serious problem is unclear in the light of similar results
of \citet{summa_16} for ray-by-ray-plus models and \citet{oconnor_16}
for fully two-dimensional two-moment transport. On the other hand, fully multi-dimensional
flux limited diffusion approaches smear out angular variations in
the radiation field too strongly \citep{ott_08_a}.

Neglecting all or part of the velocity-dependent terms
in the transport equations potentially has serious repercussions. 
Neglecting only observer correction
(Doppler shift, compression work, etc.)
as, e.g.\, in \citet{livne_04} can already have an appreciable
impact on the dynamics \citep{buras_06a,lentz_12a}.
Disregarding
even the co-advection of neutrinos with the fluid
\citep{oconnor_15,roberts_16} formally violates
the diffusion limit and effectively results in
an extra source term in the optically thick regime
due to the equilibration of matter with lagging neutrinos,
\begin{equation}
 \dot{q}_\nu \approx \rho^{-1} \mathbf{v} \cdot \nabla E_\mathrm{eq}
\end{equation}
where $E_\mathrm{eq}$ is the equilibrium neutrino energy density.
Judging from the results of \citet{oconnor_16} and \citet{roberts_16}, which
are well in line with results obtained with other codes,
the effect may not be too serious in practice, though.
It should also be noted that (semi-)stationary approximations of the
transport equation \citep{liebendoerfer_09,mueller_15a} avoid
this problem even if advection terms are not explicitly included. 

Leakage-based schemes as used, e.g., in
\citet{ott_12}, \citet{couch_14}, \citet{abdikamalov_15}, and \citet{couch_15} also manifestly fail
to reproduce the diffusion limit. Here, however, the violation of the
diffusion limit is unmistakable and can severely affect the
stratification of the gain region and, in particular, the cooling
region.  Together with \emph{ad hoc} choices for the flux factor
for calculating the heating rate,
this can result in inordinately high heating efficiencies
immediately after bounce and a completely inverted hierarchy of
neutrino mean energies. It compromises the dynamics of leakage models
to an extent that they can only be used for very qualitative studies
of the multi-D flow in the supernova core.

There is in fact no easy lesson to be learned from the pitfalls and
complications that we have outlined. In many contexts approximations
for the neutrino transport are perfectly justified for a
well-circumscribed problem, and feedback processes sometimes mitigate
the effects of simplifying assumptions. It it crucial, though, to be
aware of the impact that such approximations can potentially have, and
our (incomplete) enumeration is meant to provide some guidance in this
respect.

\begin{figure}
\begin{center}
\includegraphics[width=\linewidth]{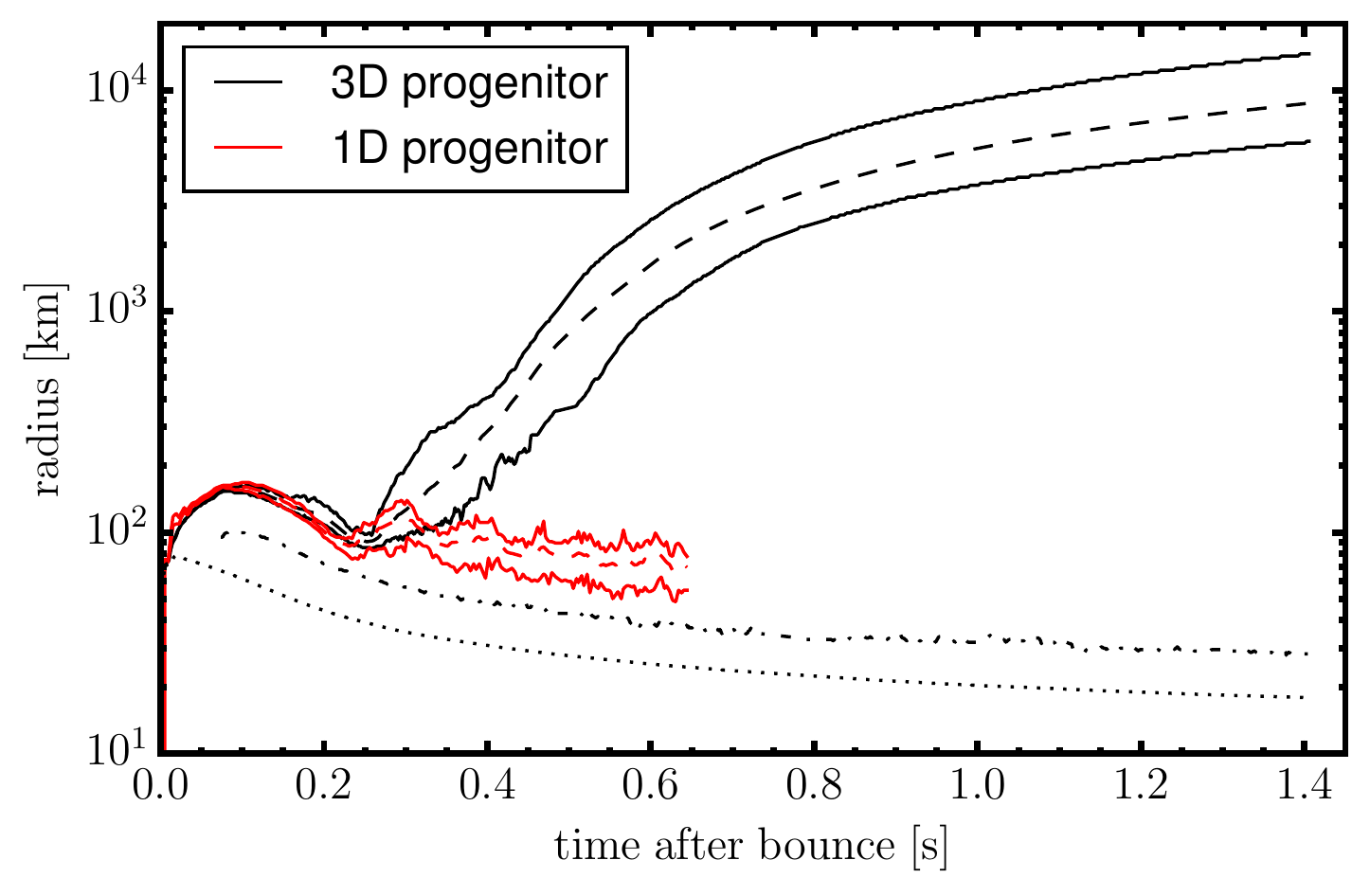}
\caption{Impact of pre-collapse asphericities on shock revival in 3D
  multi-group neutrino hydrodynamics simulations of an $18 M_\odot$
  progenitor. The plot shows the minimum, maximum (solid lines) and
  average (dashed) shock radii for a model using 3D initial conditions
  (black) from the O shell burning simulation of \citet{mueller_16b}
  and a spherically averaged version of the same progenitor (red). The
  gain radius (dash-dotted) and the proto-neutron star radius (dotted,
  defined by a fiducial density of $10^{11} \, \mathrm{g} \,
  \mathrm{cm}^{-3}$) are shown only for the model
starting from 3D initial conditions; they are virtually identical for both models.
A neutrino-driven explosion is triggered roughly $0.25 \, \mathrm{s}$
after bounce aided by the infall of the convectively perturbed oxygen
shell in the model using 3D initial conditions. The simulation
starting from the 1D progenitor model exhibits steady and strong SASI
oscillations after $0.25 \, \mathrm{s}$, but does not explode at least for
another $0.3 \, \mathrm{s}$.
\label{fig:shock_s18}}
\end{center}
\end{figure}

\section{FUTURE DIRECTIONS: MULTI-D EFFECTS IN SUPERNOVA PROGENITORS}
\label{sec:prog}
Given the sophisticated simulation methodology employed in the best
currently available supernova codes, one may be tempted to ask whether
another missing ingredient for robust neutrino-driven explosion is to
be sought elsewhere. One recent idea, first proposed by
\citet{couch_13}, focuses on the progenitor models used in supernova
simulations.  The twist consists in an extra ``forcing'' of the
non-radial motions in the gain region by large seed perturbations in
the infalling shells. Such seed perturbations will arise naturally in
active convective burning shells (O burning, and perhaps also Si burning)
that reach the shock during the first few hundred milliseconds
after bounce.

\begin{figure*}
\begin{center}
\includegraphics[width=0.48\linewidth]{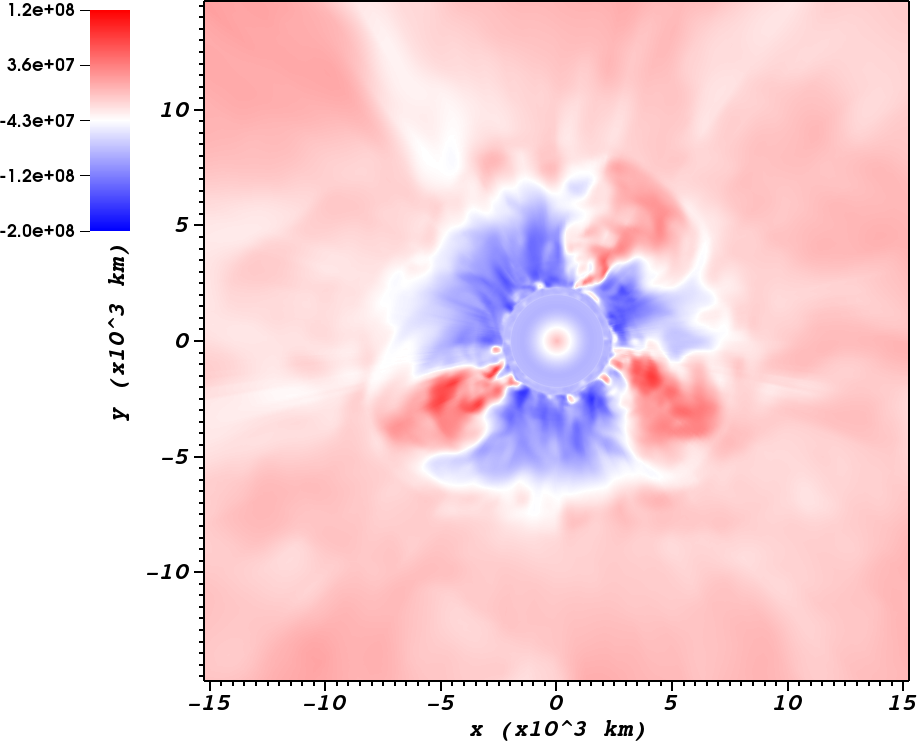}
\hfill
\includegraphics[width=0.48\linewidth]{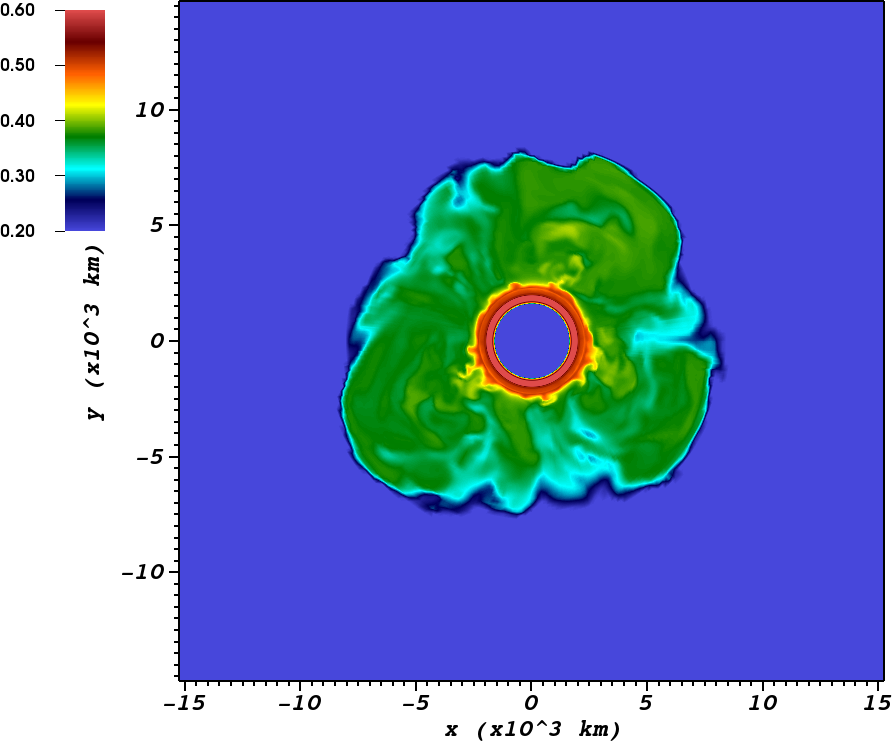}
\\
\includegraphics[width=0.48\linewidth]{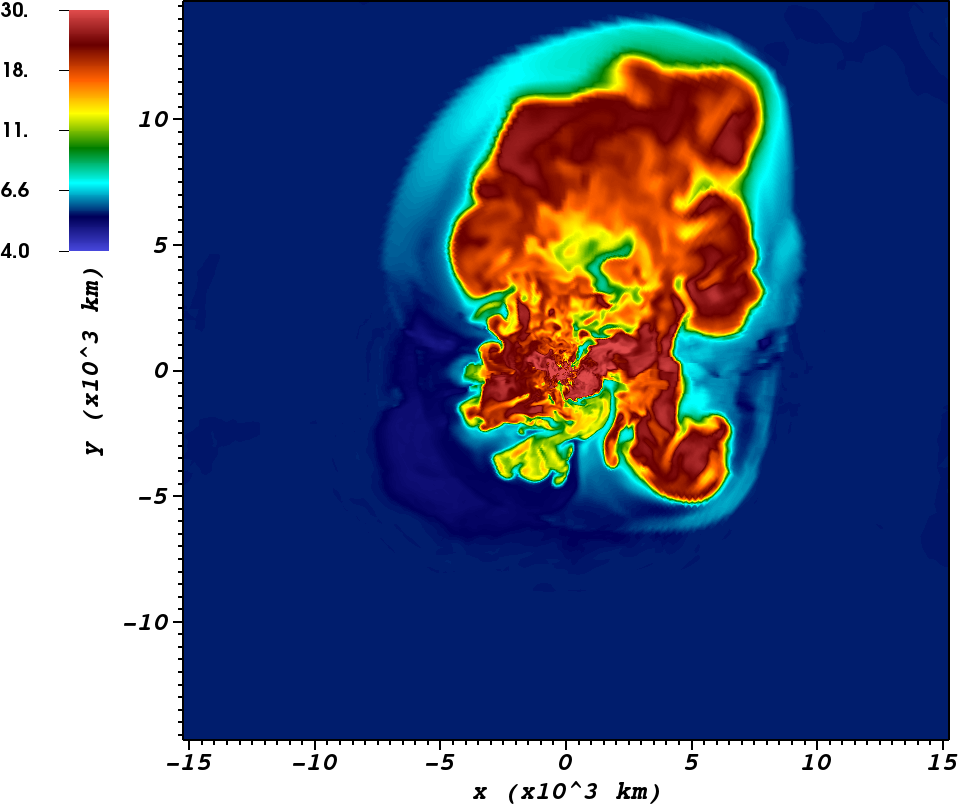}
\hfill
\includegraphics[width=0.48\linewidth]{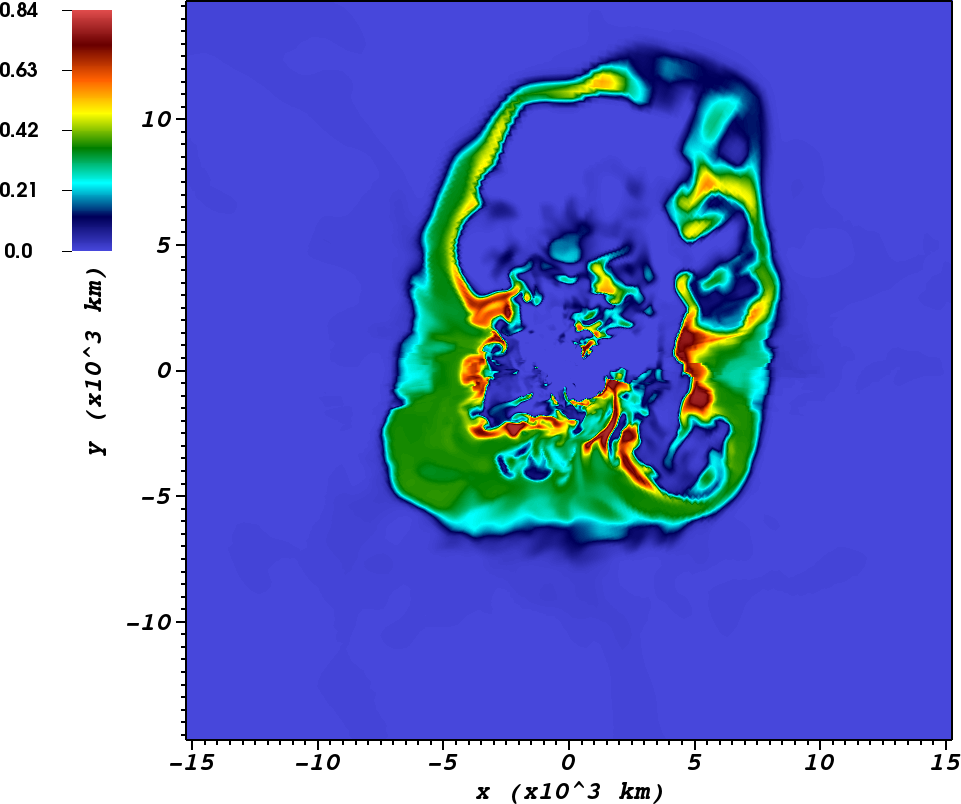}
\caption{ Top row: Radial velocity in units of $\mathrm{cm} \,
  \mathrm{s}^{-1}$ (top left) and mass fraction of Si (top right) at
  the onset of collapse in the 3D progenitor model of an $18 M_\odot$
  star of \citet{mueller_16b}.  Bottom row: Entropy in units of
  $k_\mathrm{b}/\mathrm{nucleon}$ (bottom left) and mass fraction of Si
  (bottom right) in the ensuing neutrino-driven explosion $1.43 \,
  \mathrm{s}$ after bounce from. All plots show equatorial slices from
  the 3D simulation.  It can be seen that the geometry of the initial
  conditions is still imprinted on the explosion to some extent with
  stronger shock expansion in the direction of updrafts of Si rich
  ashes in the O burning shell. This is a consequence of the forced
  deformation of the shock around the onset of the explosion.
\label{fig:s18_3d}}
\end{center}
\end{figure*}

\subsection{Role of Pre-Collapse Perturbations in the Neutrino-Driven Mechanism}
\label{sec:lcrit_pert}
In default of multi-D progenitor models, this new variation of the
neutrino-driven mechanism was initially studied by imposing large initial
perturbations by hand in leakage-based simulations
\citep{couch_13,couch_14} and multi-group neutrino hydrodynamics
simulations \citep{mueller_15a}; the earlier light-bulb based models of
\citet{fernandez_12} also touched parts of the problem.  The results
of these investigations were mixed, even though some of these
calculations employed perturbations far in excess of what estimates
based on mixing-length theory \citep{biermann_32,boehm_58} suggest:
For example, \citet{couch_13} used transverse velocity perturbations
with a peak Mach number of $\mathrm{Ma}=0.2$ in their 3D models, and
found a small beneficial effect on shock revival, which, however, was
tantamount to a change of the critical neutrino luminosity by only
$\mathord{\sim} 2\%$. The more extensive 2D parameter study of
different solenoidal and compressive velocity perturbations and
density perturbations by \citet{mueller_15a} established that both
significant perturbation velocities ($\mathrm{Ma} \gtrsim 0.1$) as
well as large-scale angular structures (angular wavenumber $\ell
\lesssim 4$) need to be present in active convective shell in order to
reduce the critical luminosity appreciably, i.e.\ by $\gtrsim 10\%$.

These parametric studies already elucidated the physical mechanism
whereby pre-collapse perturbations can facilitate shock revival.
\citet{mueller_15a} highlighted the importance both of the infall
phase as well as the interaction of the perturbations with the
shock. Linear perturbation theory shows that the initial perturbations
are amplified during collapse \citep{lai_00,takahashi_14}. This not
only involves a strong growth of transverse velocity perturbations as
$\delta v_t \propto r^{-1}$, but even more importantly a conversion of
the initially dominating solenoidal velocity perturbations with Mach
number $\mathrm{Ma}_\mathrm{conv}$ into density perturbations $\delta
\rho/\rho \approx \mathrm{Ma}$ \citep{mueller_15a} during collapse,
i.e.\ the relative density perturbations are much larger ahead of the
shock than during quasi-stationary convection, where $\delta\rho /\rho
\approx \mathrm{Ma}^2$.\footnote{I am indebted to T.~Foglizzo for
  pointing out that this conversion of velocity perturbations into
  density perturbations is another instance of advective-acoustic
  coupling \citep{foglizzo_01,foglizzo_02}, so that there is a deep,
  though not immediately obvious, connection with the physics of the
  SASI.}

Large density perturbations ahead of the shock imply a pronounced
asymmetry in the pre-shock ram pressure and deform the shock, creating
fast lateral flows as well as post-shock density and entropy
perturbations that buoyancy then converts into turbulent kinetic
energy.
The direct injection of kinetic energy due to infalling
turbulent motions may also play a role \citep{abdikamalov_16}, though
it appears to be subdominant \citep{mueller_15a,mueller_16b}.  A very
crude estimate for the generation of additional
turbulent kinetic energy due to the different processes
as well as turbulent damping in the post-shock region has
been used by \citet{mueller_16b} to estimate the reduction
of the critical luminosity as,
\begin{equation}
\label{eq:dlcrit}
(L_\nu E_\nu^2)_\mathrm{crit,pert}
\approx
(L_\nu E_\nu^2)_\mathrm{crit,3D}
\left(
1-
0.47 \frac{\mathrm{Ma_\mathrm{conv}}}{\ell \eta_\mathrm{acc} \eta_\mathrm{heat}}
\right),
\end{equation}
in terms of the pre-collapse Mach number $\mathrm{Ma}_\mathrm{conv}$
of eddies from shell burning, their typical
angular wavenumber $\ell$, and the accretion efficiency
$\eta_\mathrm{acc}=L_\nu/(GM \dot{M} r_\mathrm{gain})$ and heating
efficiency $\eta_\mathrm{heat}$ during the pre-explosion phase.  

A more rigorous understanding of the interaction between infalling
perturbations, the shock, and non-radial motions in the post-shock
region is currently emerging: \citet{abdikamalov_16} studied the
effect of upstream perturbations on the shock using the linear
interaction approximation of \citet{ribner_53} and argue, in line with
\citet{mueller_16b}, that a reduction of the critical luminosity by
$>10\%$ is plausible. Their estimate may, however, be even too
pessimistic as they neglect acoustic perturbations upstream of the
shock.  Different from \citet{abdikamalov_16}, the recent analysis of
\citet{takahashi_16} also takes into account that instabilities or
stabilisation mechanisms operate in the post-shock flow, and studied
the (linear) response of convective and SASI eigenmodes to forcing by
infalling perturbations.  A rigorous treatment along these lines that
explains the saturation of convective and SASI modes as forced
oscillators with \emph{non-linear damping} remains desirable.

\begin{figure*}
\begin{center}
\includegraphics[width=0.48\linewidth]{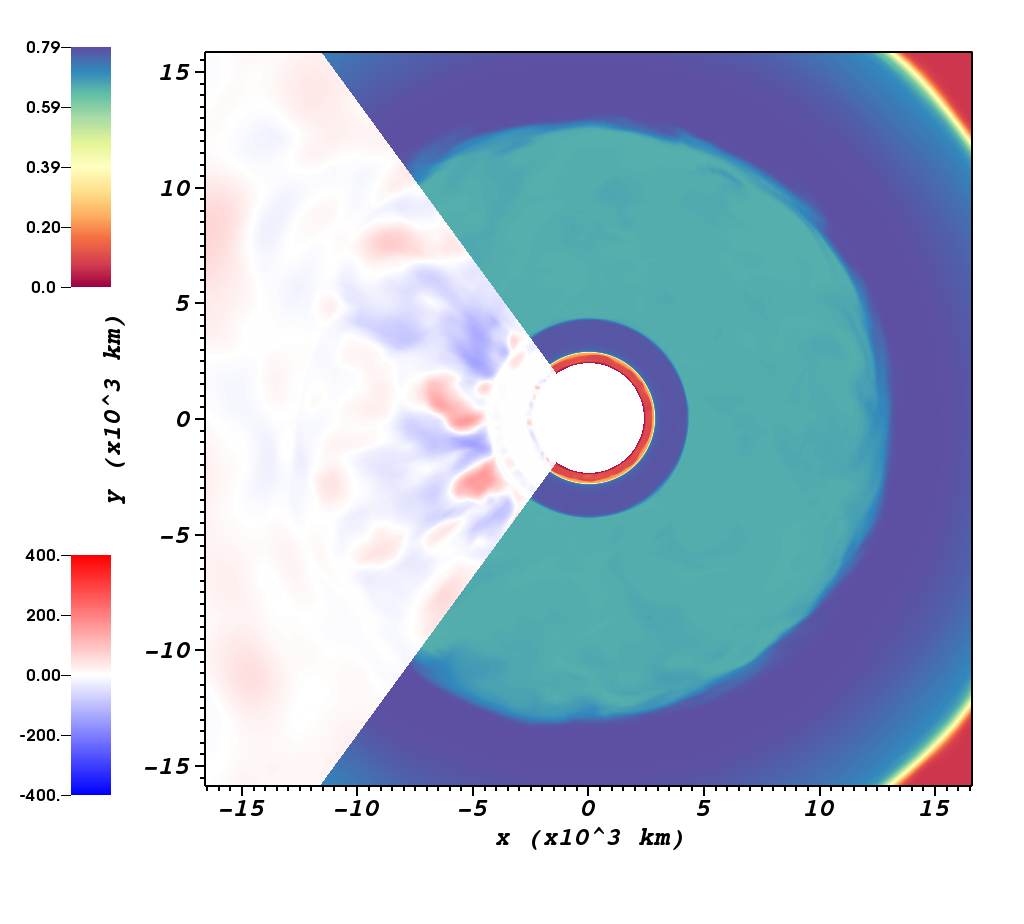}
\hfill
\includegraphics[width=0.48\linewidth]{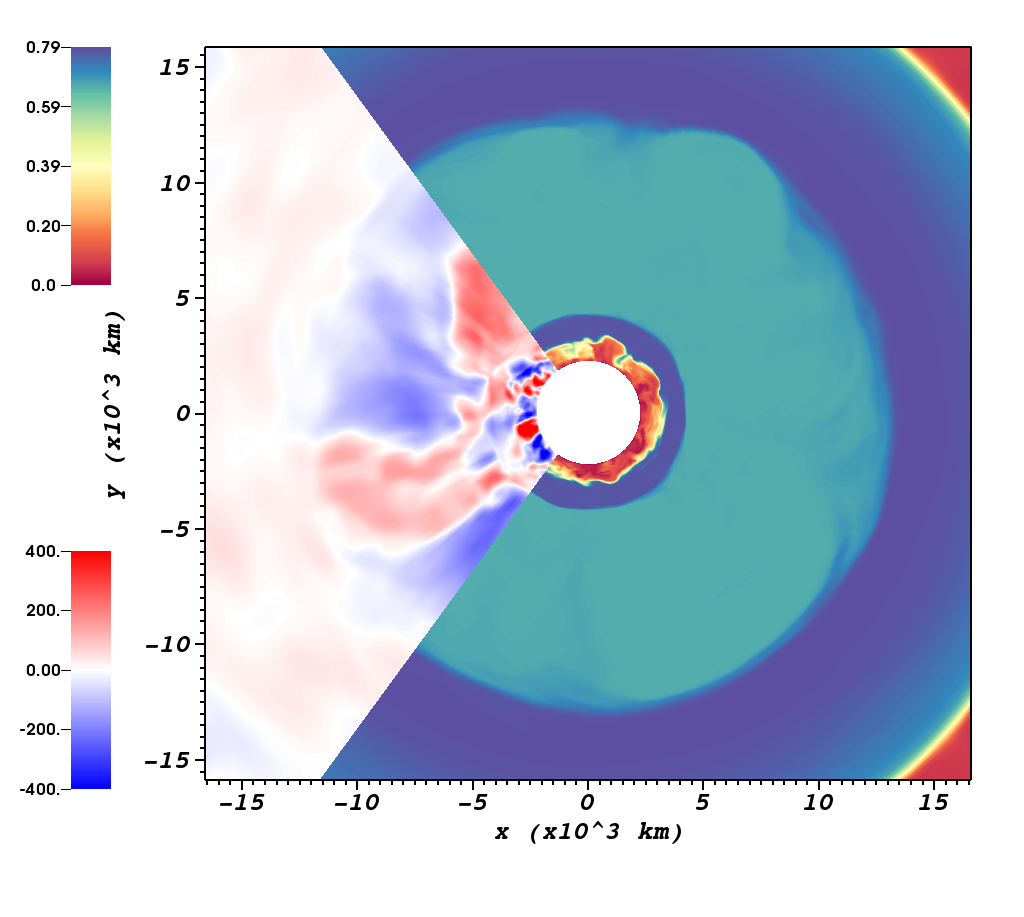}
\\
\includegraphics[width=0.48\linewidth]{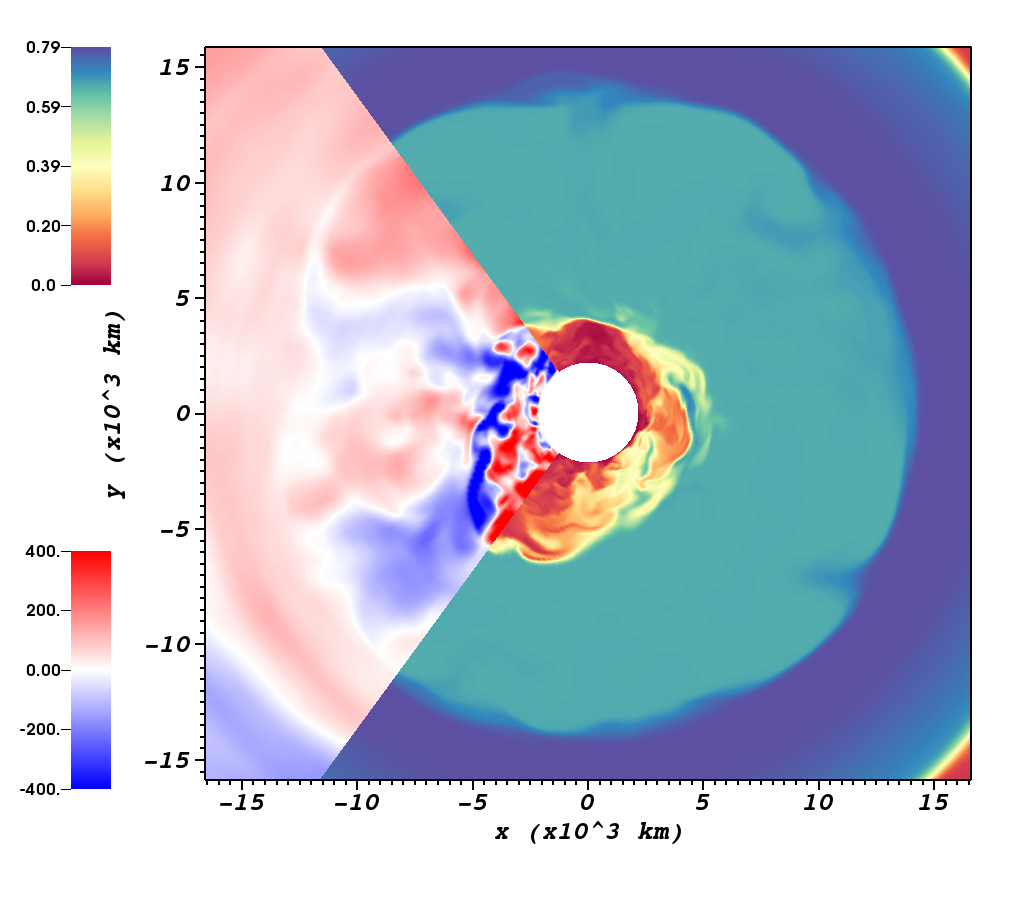}
\hfill
\includegraphics[width=0.48\linewidth]{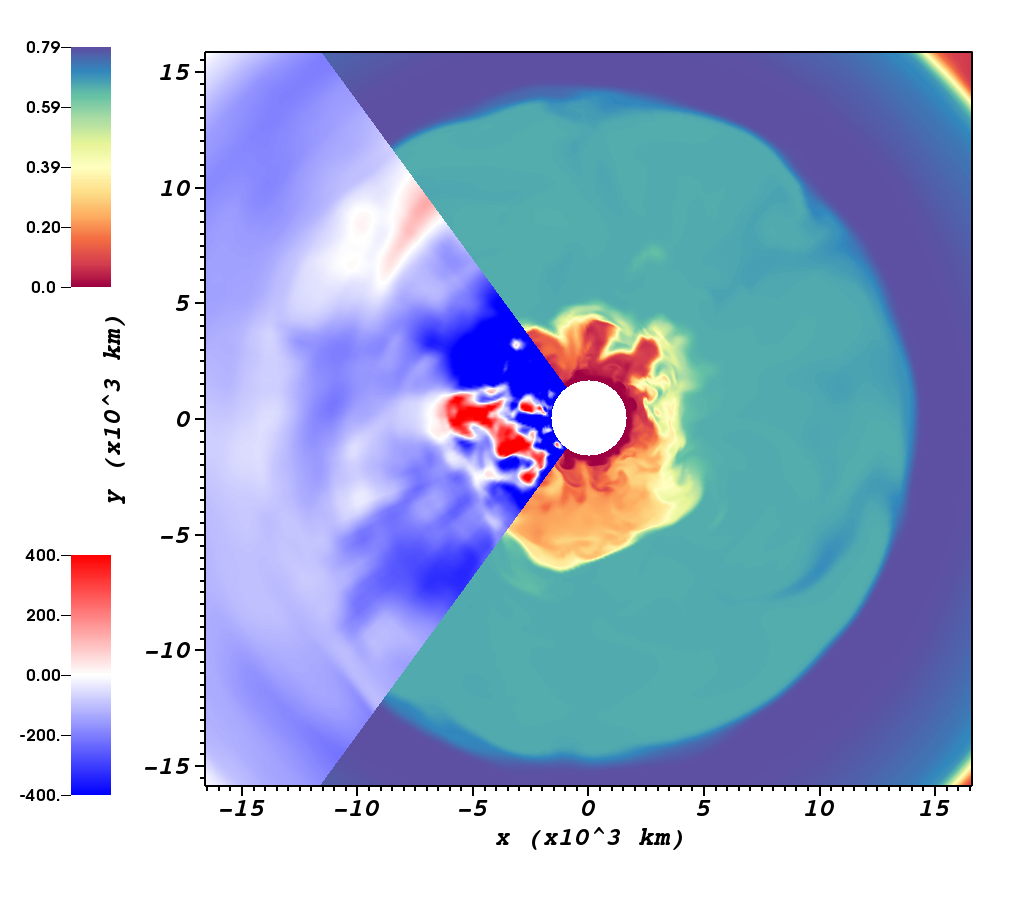}
\caption{Radial velocity in units of $\mathrm{cm} \, \mathrm{s}^{-1}$
  (shown in $90^\circ$-wedges in the left half of each plot) and mass
  fraction $X_\mathrm{O}$ of oxygen during the last minutes of shell
  burning in an $12.5 M_\odot$ progenitor.  Snapshots at at $175\,
  \mathrm{s}$ (top left), $66\, \mathrm{s}$ (top right), $24\,
  \mathrm{s}$ (bottom left) before collapse, and at the onset of
  collapse (bottom right) are shown. The residual oxygen in a
  thin, almost O-depleted shell (red) starts to burn vigorously due
to the contraction of the core (top right). As the entropy
of this shell increases and matches that of an almost
unprocessed, O-rich shell (blue) and the active Ne shell
(cyan), it expands outwards by ``encroachment'' (bottom left),
but there is insufficient time for the shells to merge
completely before collapse (bottom right). Note that this
is not a qualitatively new phenomenon in 3D; similar
events occur in 1D stellar evolution models.
\label{fig:s12_5}}
\end{center}
\end{figure*}

\subsection{The Advent of 3D Supernova Progenitor Models}
The parametric studies of \citet{couch_13,couch_14} and \citet{mueller_15a} still
hinged on uncertain assumptions about the magnitude and scale of the
seed perturbations left by O and Si shell burning.  Various pioneering
studies of advanced shell burning stages (O, Si, C burning)
\citep{arnett_94,bazan_94,bazan_98,asida_00,kuhlen_03,meakin_06,meakin_07,meakin_07_b,arnett_11,viallet_13,chatzopoulos_14}
merely indicated that convective Mach numbers of a few $10^{-2}$ and the
formation of large-scale eddies are plausible, but did not permit a clear-cut
judgement about whether pre-collapse perturbations play a dynamical role
in the neutrino-driven mechanism.

The situation has changed recently with the advent of models of
convective shell burning that have been evolved up to collapse.  The
idea here is to calculate the last few minutes prior to collapse to
obtain multi-dimensional initial conditions, while ignoring potential
long-term effects in 3D such as convective boundary mixing (which we
discuss in Section~\ref{sec:cbm}). \citet{couch_15} performed a
3D simulation of the last minutes of Si shell burning in a $15
M_\odot$ star. The simulation was limited to an octant, and nuclear
quasi-equilibrium during Si burning was only treated with a small
network. More importantly, the evolution towards collapse was
artificially accelerated by artificially increasing electron capture
rates in the iron core. As pointed out by \citet{mueller_16b}, this
can alter the shell evolution and the convective velocities
considerably. Since the shell configuration and structure at collapse
varies considerably in 1D models, such an exploratory approach is
nonetheless still justified (see below). 

\citet{mueller_16b} explored the more
generic case where Si shell burning is extinguished before collapse
and the O shell is the innermost active convective region. In their 3D
simulation of the last five minutes of O shell burning in an $18
M_\odot$ progenitor, they circumvented the aforementioned problems by
excising the non-convective Fe and Si core and contracting it in
accordance with a 1D stellar evolution model. Moreover,
\citet{mueller_16b} simulated the entire sphere using an overset
Yin-Yang grid \citep{kageyama_04,wongwathanarat_10a} as implemented
(with some improvements) in the \textsc{Prometheus} supernova code
\citep{melson_msc,melson_15a}.

The implications of these simulations for supernova modelling are
mixed. The typical convective Mach number in \citet{couch_15} was only
$\mathord{\sim}0.02$, and while they found large-scale motions, the
scale of the pre-collapse perturbations was still limited by the
restriction to octant symmetry.  Perturbations of such a magnitude are
unlikely to reduce the critical luminosity considerably
(Section~\ref{sec:lcrit_pert}).  Consequently, supernova simulations
starting from 1D and 3D initial conditions using a leakage scheme
performed by \citet{couch_15} did not show a qualitative difference;
both 1D and 3D initial conditions result in explosions, though the shock
expands slightly faster in the latter case. The use of a leakage scheme
and possible effects of stochasticity preclude definite conclusions
from these first results.

The typical convective Mach number in the $18 M_\odot$ model of
\citet{mueller_16b} is considerably larger ($\mathord{\sim}0.1$), and
their simulation also showed the emergence of a bipolar ($\ell=2$)
flow structure, which lead them to predict a relatively large
reduction of the critical luminosity by $12 \ldots 24\%$, which would
accord a decisive role to 3D initial conditions in the neutrino-driven
mechanism at least in some progenitors.  A first 3D multi-group
neutrino hydrodynamics simulation of their $18 M_\odot$ progenitor
using the \textsc{CoCoNuT-FMT} code appears to bear this out
(M\"uller et al. 2016, in preparation):
Figure~\ref{fig:shock_s18} shows the shock radius both for two
simulations using 3D and 1D initial conditions, respectively: In the former case, shock
revival occurs around $250 \, \mathrm{ms}$ after bounce thanks to the
infall of the convectively perturbed oxygen shell, whereas no
explosion develops in the reference simulation by the end of the run
more than $600 \, \mathrm{ms}$ after bounce.  An analysis of the
heating conditions indicates that the non-exploding reference model is
clearly \emph{not} a near miss at $250 \, \mathrm{ms}$. The effect of
3D initial conditions is thus unambiguously large and sufficient to
change the evolution \emph{qualitatively}.  Moreover, the model
indicates that realistic supernova explosion energies are within reach
in 3D as well: The diagnostic explosion energy reaches $5 \times 10^{50} \,
\mathrm{erg}$ and still continues to mount by the end of the
simulation $1.43 \, \mathrm{s}$ after bounce. 
 It is also interesting
to note that the initial asymmetries are clearly reflected in the
explosion geometry (Figure~\ref{fig:s18_3d}) as speculated by
\citet{arnett_11}. Incidentally, the model also shows that
the accretion of convective regions does not lead to the
formation of the ``accretion belts'' proposed
by \citet{gilkis_14} as an ingredient for their jittering-jet
mechanism.

Whether 3D initial conditions generally play an important role in the
neutrino-driven mechanism cannot be answered by studying just two
progenitors, aside from the fact that the models of \citet{couch_15}
and \citet{mueller_16b} still suffer from limitations.  The properties
(width, nuclear energy generation rate) and the configuration of
convective burning shells at collapse varies tremendously across
different progenitors in 1D stellar evolution models as, e.g., the
Kippenhahn diagrams in the literature indicate
\citep{heger_00,chieffi_13,sukhbold_14,cristini_16} indicate. The
interplay of convective burning, neutrino cooling, and the
contraction/re-expansion of the core and the shells sometimes leave
inversions in the temperature stratification and a complicating
layering of material at different nuclear processing stages.  For this
reason, 1D stellar evolution models sometimes show a highly dynamic
behaviour immediately prior to collapse with shells of incompletely
burnt material flaring up below the innermost active shell.  This is
illustrated by follow-up work to \citet{mueller_16b} shown in
Figure~\ref{fig:s12_5}, where a partially processed layer with unburnt
O becomes convective shortly before collapse due to
violent burning and is about to merge
with the overlying O/Ne shell before collapse intervenes.

The diverse shell configurations in supernova progenitors need to be
thoroughly explored in 3D before a general verdict on the efficacy of
convective seed perturbations in aiding shock revival can be given.
Since the bulk properties of the flow (typical velocity, eddy scales)
in the \emph{interior} of the convective shells are apparently well
captured by mixing-length theory \citep{arnett_09,mueller_16b}, the
convective Mach numbers and eddy scales predicted from 1D stellar
evolution models can provide guidance for exploring interesting spots
in parameter space.

\subsection{Convective Boundary Mixing -- How Uncertain
is the Structure of Supernova Progenitors?}
\label{sec:cbm}
In what we discussed so far, we have considered multi-D effects in
advanced convective burning stages merely because of their role in
determining the initial conditions for stellar collapse. They could
also have an important effect on the secular evolution of massive
stars long before the supernova explosion, and thereby change critical
structural properties of the progenitors, such as the compactness
parameter \citep{oconnor_11}. While mixing-length theory
\citep{biermann_32,boehm_58} may adequately describe the mixing in the
interior of convective zones,\footnote{The story may
be different for angular momentum
transport in convective zones, which deserves to revisited
(see  \citealt{chatzopoulos_16} for a current study
in the context of Si and O shell burning).} the mixing across convective
boundaries is less well understood, and may play an important role in
determining the pre-collapse structure of massive stars along with
other non-convective processes
\citep[e.g.][]{heger_00,maeder_04,heger_05,young_05,talon_05,cantiello_14}
for mixing and angular momentum transport.  That some mixing beyond
the formally unstable regions needs to be included has long been known
\citep{kippenhahn}.  Phenomenological recipes for this include
extending the mixed region by a fraction of the local pressure scale
height, or  adding diffusive mixing in
the formally stable regions with a calibrated functional dependence on
the distance to the boundary \citep{freytag_96,herwig_97}.

The dominant mechanism for convective boundary mixing during advanced
burning stages is entrainment \citep{fernando_91,meakin_07,viallet_15}
due to the growth of the Kelvin-Helmholtz or Holmb\"oe instability at
the shell interfaces. For interfaces with a discontinuous density
jump as often encountered in the interiors of evolved massive stars,
the relevant dimensionless number for such
shear-driven instabilities
is the bulk Richardson number $\mathrm{Ri}_\mathrm{B}$.  For
entrainment driven by turbulent convection, one has
\begin{equation}
\mathrm{Ri}_\mathrm{B}=\frac{g l\, \delta \rho/\rho }{v_\mathrm{conv}^2},
\end{equation}
 in terms of the local gravitational acceleration $g$, the density
 contrast $\delta\rho /\rho$ at the interface, the typical convective velocity
 $v_\mathrm{conv}$ in the convective region, and the integral scale $l$
 of the convective eddies. Equating $l$ with the pressure scale height
$l=P/\rho g$ allows us to re-express $  \mathrm{Ri}_\mathrm{B}$ in
terms of the convective Mach number $ \mathrm{Ma}_\mathrm{conv}$
and the adiabatic exponent $\gamma$,
\begin{equation}
  \mathrm{Ri}_\mathrm{B}=
  \frac{\delta \rho}{\rho} \frac{g l}{v_\mathrm{conv}^2}
=
  \frac{\delta \rho}{\rho} \frac{P}{\rho v_\mathrm{conv}^2}
=
  \frac{\delta \rho}{\rho} \frac{1}{\gamma \mathrm{Ma}_\mathrm{conv}^2}.
\end{equation}
Deep in the stellar core, $\mathrm{Ma}_\mathrm{conv}$ is typically small during
most evolutionary phases, and $  \mathrm{Ri}_\mathrm{B}$ is large so that
the convective boundaries are usually very ``stiff'' \citep{cristini_16}.

Various power laws for the entrainment rate have been proposed in the
general fluid dynamics literature \citep{fernando_91,strang_01} and
astrophysical studies \citep{meakin_07} of interfacial mixing  driven
by turbulent convection on one side of the interface.  In the
astrophysical context, it is convenient to translate these into a
power law for the mass flux $\dot{M}_\mathrm{entr}$ of entrained
material into the convective region,
\begin{equation}
\label{eq:mentr}
  \dot{M}_\mathrm{entr}
  =4 \pi r^2 \rho v_\mathrm{conv} A \, \mathrm{Ri}_\mathrm{B}^{-n},
\end{equation}
with a proportionality constant $A$ and a power-law exponent $n$.
Here $\rho$ is the density on the convective side of the interface.

A number of laboratory studies \citep{fernando_91,strang_01} and
astrophysical simulations \citep{meakin_07,mueller_16b} suggest values
of $A\sim 0.1$ and $n=1$. This can be understood heuristically
by assuming that layer of width
$\delta l \sim A v_\mathrm{conv}^2/(g \, \delta\rho/\rho)$
always remains well mixed,\footnote{The
width of this region will be determined by the criterion
that the gradient Richardson number is about $1/4$.} and that
a fraction $\delta l/l$ of the mass flux $\dot{M}_\mathrm{down}
=2 \pi r^2 \rho v_\mathrm{conv}$ in the convective downdrafts
comes from this mixed layer.

This estimate is essentially equivalent to another one proposed in a
slightly different context (ingestion of unburnt He during core-He
burning; \citealp{constantino_16}) by \citet{spruit_15}, who related
the ingestion (or entrainment) rate into a convective zone to the
convective luminosity $L_\mathrm{conv}$. Spruit's argument
can be interpreted as one based on energy conservation; work is
needed to pull material with positive buoyancy from an outer
shell down into a deeper one, and the energy that is tapped
for this purpose comes from convective motions.
Since $L_\mathrm{conv} \sim 4\pi r^2 \rho v_\mathrm{conv}^3$,
we can write Equation~(\ref{eq:mentr}) as
\begin{equation}
  \label{eq:mentr2}
  \dot{M}_\mathrm{entr}
  =A \times \frac{4 \pi r^2 \rho v_\mathrm{conv}^3 }{g l\,\delta \rho /\rho}
\approx A \times \frac{L_\mathrm{conv}}{g l \, \delta \rho /\rho },
\end{equation}
which directly relates the entrainment rate to the ratio of
$L_\mathrm{conv}$ and the potential energy of material with positive
buoyancy after downward mixing over an eddy scale $l$.  The
entrainment law (\ref{eq:mentr}), the argument of \citet{spruit_15},
and the proportionality of the entrainment rate with $L_\mathrm{conv}$
found in the recent work of \citet{jones_16b} on entrainment in
highly-resolved idealised 3D simulation of O shell burning appear to
be different sides of the same coin.

\subsection{Long-Term Effects of Entrainment on the Shell Structure?} 
\label{sec:long_term}
How much will entrainment affect the shell structure of massive stars
in the long term? First numerical experiments based on
the entrainment law of \citet{meakin_07} were performed by
\citet{staritsin_13} for massive stars on the main sequence
\footnote{It is doubtful whether entrainment operates
efficiently for core H burning, though. Here diffusivity
effects are not negligible for convective boundary mixing,
which is thus likely to take on a different character \citep{viallet_15}.} and did not reveal
dramatic differences in the size of the convective cores
compared to more familiar, calibrated recipes for core overshooting.

Taking Equation~(\ref{eq:mentr2}) at face value allows some
interesting speculations about the situation during advanced burning
stages. Since the convective motions ultimately feed on the energy
generated by nuclear burning $E_\mathrm{burn}$, we can formulate a
time-integrated version of Equation~(\ref{eq:mentr2}) for the
entrained mass $\Delta M_\mathrm{entr}$ over the life time of a
convective shell,
\begin{eqnarray}
 \frac{GM}{r} \frac{\delta \rho}{\rho } \Delta M_\mathrm{entr}
&\lesssim &
A E_\mathrm{burn}, \\
 \frac{GM}{r} \frac{\delta \rho}{\rho } \Delta M_\mathrm{entr}
&\lesssim &
A M_\mathrm{shell} \Delta Q,
\end{eqnarray}
where $M_\mathrm{shell}$ is the (final)
mass of the shell, and $\Delta Q$ is the nuclear energy release per unit mass. With
$GM/r \sim 2 e_\mathrm{int}$ in stellar interiors, we can estimate $\Delta M_\mathrm{entr}$
in terms $\Delta Q$ and the internal energy $e_\mathrm{int}$ at which the burning occurs,\footnote{
$e_\mathrm{int}$ at the shell boundary may be the more relevant scale, but the convective
luminosity typically decreases even more steeply with $r$ than $e_\mathrm{int}$, so our
estimate is on the safe side for formulating an upper limit.}
\begin{equation}
\Delta M_\mathrm{entr}
\lesssim A M_\mathrm{shell} \left(\frac{\delta \rho }{\rho}\right)^{-1} \frac{\Delta Q}{2 e_\mathrm{int} }.
\end{equation}
For O burning at $\sim 2 \times 10^{9} \, \mathrm{K}$ and with $\Delta
Q \approx 0.5 \, \mathrm{MeV}/ \mathrm{nucleon}$, the factor $\Delta
Q/(2 e_\mathrm{int})$ is of order unity.  Typically, the density
contrast $\delta \rho/\rho$ between adjacent shells is also not too
far below unity.  Since $A\approx 0.1$, this suggests that the shell
growth due to entrainment comes up to at most a few tens of percent
during O shell burning unless $\delta \rho/\rho$ is rather small to
begin with. Thus, a result of entrainment might be that convective
zones may swallow thin, unburnt shells with a small density contrast
before bounce, whereas the large entropy jumps between the major
shells are maintained and even enhanced as a result of this
cannibalisation.

For C burning, the long-term effect of entrainment could be somewhat
larger than for O burning due to the lower temperature
threshold and the higher ratio $\Delta Q/2 e_\mathrm{int}$; for
Si burning, the effect should be smaller. During earlier phases our estimates break down because the
convective flux carries only a small fraction of the energy generation
by nuclear burning. If this is taken into account,
the additional growth of convective regions due to entrainment 
is again of a modest scale \citep{spruit_15}.

\subsection{Caveats}
The  estimates for the long-term effect of entrainment on
the growth of convective regions in Section~\ref{sec:long_term}
are to be taken with caution, however. They are not only
crude, time-integrated zeroth-order estimates; the
entrainment law~(\ref{eq:mentr2}) is by no means set
in stone.  Current astrophysical 3D simulations
only probe a limited range in the critical parameter
$ \mathrm{Ri}_\mathrm{B}$, and tend to suffer from
insufficient resolution for high $ \mathrm{Ri}_\mathrm{B}$,
as shear instabilities develop on smaller and smaller scales.

As a result, it cannot be excluded that the entrainment
law~(\ref{eq:mentr}) transitions to a steeper slope in the
astrophysically relevant regime of high $
\mathrm{Ri}_\mathrm{B}$. Experiments also compete with the
difficulties of a limited dynamic range in Reynolds, Prandtl, and
P\'eclet number, and remain inconclusive about the regime of high
$\mathrm{Ri}_\mathrm{B}$ that obtains in stellar interiors.  Power-law
exponents larger than $n=1$ (up to $n=7/4$) have also been reported in
this regime as alternatives to $n=1$
\citep{fernando_91,strang_01,fedorovich_04}. A power-law exponent
$n>1$ would imply a strong suppression of entrainment in stellar
interiors under most circumstances, and the long-term effect of
entrainment would be negligible.  Moreover, magnetic fields will
affect the shear-driven instabilities responsible for convective
boundary mixing \citep{brueggen_01}.

Finally, most of the current 3D simulations of convective boundary mixing suffer
from another potential problem; the balance between nuclear
energy generation and neutrino cooling that obtains
during quasi-stationary shell burning stages is typically violated,
or neutrino cooling is not modelled at all. \citet{jones_16b}
pointed out that this may be problematic if neutrino cooling
decelerates the buoyant convective plumes and reduces
the shear velocity at the interfacial boundary. Only sufficiently
long simulations will be able clarify whether the strong
entrainment seen in some numerical simulations is robust or
(partly) specific to a transient adjustment phase.

Thus, it remains to be seen whether convective boundary mixing has
significant effects on the structure of supernova progenitors.
Even if it does, it is not clear whether it will qualitatively
affect the landscape of supernova progenitors. The general
picture of the evolution of massive stars may stay well within
the bounds of the variations that have been explored
already, albeit in a more parametric way \citep[see, e.g.,][]{sukhbold_14}.

\section{CONCLUSIONS}
It is evident that our understanding of the supernova explosion
mechanism has progressed considerably over the last few years. While
simulations of core-collapse supernovae have yet to demonstrate that
they can correctly reproduce and explain the whole range explosions
that is observed in nature, there are plenty of ideas for solving the
remaining problems. Some important milestones from the last few
years have been discussed in this paper, and can be summarised
as follows:
\begin{itemize}
\item
ECSN-like explosions of supernova progenitors with the lowest masses ($8\ldots
10 M_\odot$) can be modelled successfully both in 2D and in
3D. Regardless of the precise evolutionary channel from which they
originate, supernovae from the transition region between the super-AGB
star channel and classical iron-core collapse supernovae share similar
characteristics, i.e.\ low explosion energies of $\mathord{\sim}10^{50}
\, \mathrm{erg}$ and small nickel masses of a few $10^{-3} M_\odot$.
Due to the ejection of slightly neutron-rich material in the early
ejecta, they are an interesting source site for the production of the lighter
neutron-rich trans-iron elements (Sr, Y, Zr), and are potentially
even a site for a weak r-process up to Ag and Pd \citep{wanajo_11}.
An unambiguous identification of ECSN-like explosions among
observed transients is still pending, however, although there
are various candidate events.
\item
Though it has yet to be demonstrated that the neutrino-driven
explosion mechanism can robustly account for the explosions of more
massive progenitors, first successful 3D models employing
multi-group neutrino transport have recently become available.
The reluctance of the first 3D models to develop explosions
due to the different nature of turbulence in 3D proves to be no
insurmountable setback; and even the unsuccessful
3D models computed so far appear to be close to explosion.
\item Some of the recent 2D models produced by different groups
  \citep{summa_16,oconnor_16} show similar results, which inspires
  some confidence that the simulations are now at a stage where
  modelling uncertainties due to different numerical methodologies are
  under reasonable control, though they have not been completely
  eliminated yet.  We have addressed some of the sensitivities to the
  modelling assumption in this paper, including possible effects of
  numerical resolution as well as various aspects of the neutrino
  transport treatment.
\item Recent studies have helped to unravel how the interplay
between neutrino heating and hydrodynamic instabilities works
quantitatively, and they have clarified why neutrino-driven
mechanism can be obtained with a considerably smaller driving luminosity
in multi-D. 
\item There is a number of ideas about missing physics that
could make the neutrino-driven mechanism robust for a wider
range of progenitors. These include rapid rotation (\citealp{nakamura_14,janka_16}; though
stellar evolution makes this unlikely as a generic explanation),
changes in the neutrino opacities \citep{melson_15b},
and a stronger forcing of non-radial instabilities
due to seed perturbations from convective shell burning
\citep{couch_13,couch_15,mueller_15a,mueller_16b}.
\item 3D initial conditions for supernova simulations have now become
  available \citep{couch_15,mueller_16b}, and promise to play a
  significant and beneficial role in the explosion mechanism. A first
  3D multi-group simulation starting from a 3D initial model of an $18
  M_\odot$ progenitor has been presented in this review. The model
has already reached an explosion energy of $5 \times 10^{50} \, \mathrm{erg}$,
and suggests that the observed range of explosion energies may be
within reach of 3D simulations.
\item Nonetheless, the study of 3D effects in supernova progenitors is yet
in its infancy. A thorough exploration of the parameter space
is required in order to judge whether they are generically important
for our understanding of supernova explosions. This is not
only true with regard to the 3D pre-collapse
perturbations from shell burning that are crucial to the
``perturbation-aided'' neutrino-driven mechanism. The role of
convective boundary mixing on the structure of supernova
progenitors also deserves to be explored.
\end{itemize}
Many of these developments are encouraging, though there are also
hints of new uncertainties that may plague supernova theory in the
future. Whether the new ideas of recent years will prove sufficient to
explain shock revival in core-collapse supernovae remains to be
seen. The perspectives are certainly good, but obviously a lot more
remains to be done before simulations and theory can fully explain the
diversity of core-collapse events in nature. There is no need to
fear a shortage of fruitful scientific problems concerning the
explosions of massive stars.

{
\begin{acknowledgements}
The author acknowledges fruitful discussions with R.~Bollig,
A.~Burrows, S.~Couch, E.~Lentz, Th.~Foglizzo, A.~Heger, F.~Herwig,
W.~R.~Hix, H.-Th.~Janka, S.~Jones, T.~Melson, R.~Kotak, J.~Murphy,
K.~Nomoto, E.~O'Connor, L.~Roberts, S.~Smartt, H.~Spruit, and
M.~Viallet.  Particular thanks go to A.~Heger, S.~Jones, and K.~Nomoto
for providing density profiles of ECSN-like progenitors for
Figure~\ref{fig:threshold}, to H.-Th.~Janka for critical reading,
and to T.~Melson and M.~Viallet for long-term assistance with
the development of the \textsc{Prometheus} code.
Part of this work has been supported by the Australian Research
Council through a Discovery Early Career Researcher Award (grant
DE150101145).  This research was undertaken with the assistance of
resources from the National Computational Infrastructure (NCI), which
is supported by the Australian Government. This work was also
supported by resources provided by the Pawsey Supercomputing Centre
with funding from the Australian Government and the Government of
Western Australia, and by the National Science Foundation under Grant
No. PHY-1430152 (JINA Center for the Evolution of the Elements).
Computations were performed on the systems \emph{raijin} (NCI) and
\emph{Magnus} (Pawsey), and also on the IBM iDataPlex system
\emph{hydra} at the Rechenzentrum of the Max-Planck Society (RZG) and
at the Minnesota Supercomputing Institute.
\end{acknowledgements}
}

\appendix
\section{The Density Gradient in the Post-Shock Region}
\label{sec:app_gradient}
Neglecting quadratic terms in the velocity and neglecting
the self-gravity of the material in the gain region, one can write
the momentum and energy equation for quasi-stationary accretion
onto the proto-neutron star in the post-shock region as,
\begin{eqnarray}
\frac{1}{\rho}\frac{\pd P}{\pd r}
&=&-\frac{GM}{r^2}, \\
\label{eq:a2}
\frac{\pd}{\pd r}\left(h-\frac{GM}{r}\right)
&=&\frac{\dot{q}_\nu}{v_r},
\end{eqnarray}
in terms of the pressure $P$, the density $\rho$,
the proto-neutron star mass  $M$, the
enthalpy $h$, the mass-specific net neutrino heating
rate $\dot{q}_\nu$, and the radial velocity $v_r$.
For a radiation-dominated gas, one has $h\approx 4P/\rho$,
which implies,
\begin{equation}
\label{eq:a3}
\frac{1}{4}
\frac{\pd h}{\pd r}
+\frac{h}{4}\frac{\pd \ln \rho}{\pd r}
=
-\frac{GM}{r^2},
\end{equation}
and by taking $\pd h/\pd r$ from Equation~(\ref{eq:a2}),
\begin{equation}
  \label{eq:a4}
\frac{\dot{q}_\nu}{4v_r}+
\frac{h}{4}\frac{\pd \ln \rho}{\pd r}
=
-\frac{3GM}{4 r^2 }.
\end{equation}
Solving for the local power-law slope
$\alpha=\pd \ln \rho /\pd \ln r$
of the density yields,
\begin{equation}
\alpha
=
-\frac{3GM}{r h}-\frac{r \dot{q}_\nu}{v_r h}.
\end{equation}
Since $\dot{q}_\nu>0$ and $v_r<0$ in the gain
region before shock revival, this implies a power-law
slope $\alpha$ that is no steeper than,
\begin{equation}
\alpha
\geq
-\frac{3GM}{r h}.
\end{equation}

\bibliographystyle{pasa-mnras}
\bibliography{paper}

\end{document}